\documentclass[twocolumn, twocolappendix]{aastex631}

\received{2023 November 17}
\revised{2024 January 15}
\accepted{2024 January 26}

\submitjournal{PASP}

\shorttitle{High-precision astrometry and photometry with {\jwst}/MIRI}

\shortauthors{Libralato et al.}

\usepackage{xspace}
\usepackage{amsmath}
\usepackage{rotating}
\usepackage{threeparttable}
\usepackage{enumerate}
\usepackage{color}

\newcommand{\hstfull}{\textit{Hubble Space Telescope}\xspace}
\newcommand{\hst}{\textit{HST}\xspace}

\newcommand{\jwst}{\textit{JWST}\xspace}
\newcommand{\gaia}{\textit{Gaia}\xspace}
\newcommand{\qfit}{\texttt{QFIT}\xspace}
\newcommand{\radxs}{\texttt{RADXS}\xspace}

\begin{document}

\title{High-precision astrometry and photometry with the {\jwst}/MIRI imager}

\correspondingauthor{Mattia Libralato}
\email{libra@stsci.edu}

\author[0000-0001-9673-7397]{Mattia Libralato}
\affil{AURA for the European Space Agency (ESA), Space Telescope Science Institute, 3700 San Martin Drive, Baltimore, MD 21218, USA}

\author[0000-0003-2820-1077]{Ioannis Argyriou}
\affil{Institute of Astronomy, KU Leuven, Celestijnenlaan 200D, 3001 Leuven, Belgium}

\author[0000-0003-0589-5969]{Dan Dicken}
\affiliation{UK Astronomy Technology Centre, Royal Observatory Edinburgh, Blackford Hill, Edinburgh EH9 3HJ, UK}

\author[0000-0003-4801-0489]{Macarena Garc\'{\i}a Mar\'{\i}n}
\affiliation{European Space Agency, Space Telescope Science Institute, 3700 San Martin Drive, Baltimore, MD, 21218, USA}

\author[0000-0002-2421-1350]{Pierre Guillard}
\affiliation{Sorbonne Universit\'{e}, CNRS, UMR 7095, Institut d'Astrophysique de Paris, 98bis bd Arago, 75014 Paris, France}
\affiliation{Institut Universitaire de France, Minist{\`e}re de l'Enseignement Sup{\'e}rieur et de la Recherche, 1 rue Descartes, 75231 Paris Cedex 05, France}

\author[0000-0003-4653-6161]{Dean C. Hines}
\affiliation{Space Telescope Science Institute, 3700 San Martin Drive, Baltimore, MD, 21218, USA}

\author[0000-0001-6872-2358]{Patrick J. Kavanagh}
\affiliation{Department of Experimental Physics, Maynooth University, Maynooth, Co.
Kildare, Ireland}

\author[0000-0002-7612-0469]{Sarah Kendrew}
\affiliation{European Space Agency, Space Telescope Science Institute, 3700 San Martin Drive, Baltimore, MD, 21218, USA}

\author[0000-0002-9402-186X]{David R. Law}
\affiliation{Space Telescope Science Institute, 3700 San Martin Drive, Baltimore, MD, 21218, USA}

\author[0000-0002-6296-8960]{Alberto Noriega-Crespo}
\affiliation{Space Telescope Science Institute, 3700 San Martin Drive, Baltimore, MD, 21218, USA}

\author[0000-0002-7093-1877]{Javier \'Alvarez-M\'arquez}
\affiliation{Centro de Astrobiolog\'ia (CAB, CSIC-INTA), Carretera de Ajalvir, E-28850 Torrej\'on de Ardoz, Madrid, Spain}

\begin{abstract}
Astrometry is one of the main pillars of astronomy, and one of its oldest branches. Over the years, an increasing number of astrometric works by means of \hstfull (\hst) data have revolutionized our understanding of various phenomena. With the launch of \jwst, it becomes almost instinctive to want to replicate or improve these results with data taken with the newest, state-of-the-art, space-based telescope. In this regard, the initial focus of the community has been on the Near-Infrared (NIR) detectors on board of \jwst because of their high spatial resolution. This paper begins the effort to capture and apply what has been learned from \hst to the Mid-InfraRed Instrument (MIRI) of \jwst by developing the tools to obtain high-precision astrometry and photometry with its imager. We describe in detail how to create accurate effective point-spread-function (ePSF) models and geometric-distortion corrections, analyze their temporal stability, and test their quality to the extent of what is currently possible with the available data in the \jwst MAST archive. We show that careful data reduction provides deep insight on the performance and intricacies of the MIRI imager, and of \jwst in general. In an effort to help the community to devise new observing programs, we make our ePSF models and geometric-distortion corrections publicly available.
\end{abstract}


\section{Introduction}\label{sec:intro}

The first six months after the launch of \jwst \citep{2023GardnerJWST} focused on its Commissioning \citep{2023RigbyJWSTCom} by performing a series of on-sky and internal source calibration observations and data analyses aimed at understanding the actual performance of the telescope and its instruments. The characterization of each subsystem, together with why and by how much its capabilities differed from pre-launch expectations, was aimed at providing to the community the necessary calibrations to make scientific analyses of \jwst data possible from the start. The Commissioning activities covered as many aspects as possible, but due to the limited time allocated to Commissioning, the more detailed characterization of the instrument performance, and development of sophisticated tools for their analysis, was postponed to the Cycle 1.

An example is the lack of tools to obtain high-precision astrometry with \jwst's imagers. The experience with the \hstfull (\hst) has shown that it is possible to achieve exquisite astrometric (and photometric) accuracies and precisions thanks to accurate effective point-spread-function (ePSF) models\footnote{Following the convention in \citet{2000AKWFPC2}, the ePSF is defined as the convolution of the PSF due to the telescope optics with the spatial sensitivity function of a pixel.} and geometric-distortion (GD) corrections. While ePSF models were not released after Commissioning, GD corrections are available in the \jwst Calibration Reference Data System (CRDS). However, these GD corrections were designed to meet the mission requirements (GD uncertainty for all detectors to be below 5 mas per coordinate; see Anderson 2016\footnote{Document \href{https://www.stsci.edu/files/live/sites/www/files/home/jwst/documentation/technical-documents/_documents/JWST-STScI-005361.pdf}{JWST-STScI-005361} ``Verification of Plan to Solve
for the Distortion Solution''.}), leaving further refinements that can be critical for some scientific investigations to the Cycle-1 Calibration process.

For the cameras onboard of the \hst, it took a few years to have the first tools to achieve high-precision astrometry \citep{2000AKWFPC2}. For \jwst, the community has already provided initial sets of ePSFs and GD corrections for the near-infrared (NIR) detectors of the Near InfaRed Camera \citep[NIRCam,][]{2022NardielloNIRCam,2023GriggioNIRCam} and the Near Infrared Imager and Slitless Spectrograph \citep[NIRISS,][]{2023LibralatoNIRISS}. Only the Mid-InfraRed (MIR) capabilities of \jwst have not been addressed yet: our paper strives to begin filling this gap in the astrometric cause of \jwst.

This manuscript describes the efforts of the Mid-InfraRed Instrument \citep[MIRI,][]{2015RiekeMIRI,2023WrightMIRI} team, and in particular of its imager working group, to provide a set of ePSF models and GD corrections for the MIRI imager, together with the description of how they were made. Our analyses show that high-precision astrometry and photometry are also within the capabilities of the MIRI imager.

The paper is organized as follows: Sect.~\ref{sec:data} presents the data used in this work; Sects.~\ref{sec:psf} and \ref{sec:gd} describe how ePSF models and GD corrections were made; Sect.~\ref{sec:desc} provides an overview of the tools we release; and finally Sect.~\ref{sec:science} showcase our work with a simple scientific application.

\section{Instrument Description and Data sets}\label{sec:data}

The MIRI imager (MIRIM) uses a Si:As blocked impurity band conduction detector with 1032$\times$1024 pixel$^2$. Not all pixels are exposed to the incoming light \citep{2015ResslerMIRI,2023MorrisonMIRI}. What is commonly addressed as the imager is the largest part on the right side of the detector with a Field of View (FoV) of about 74$\times$113 arcsec$^2$ \citep[nominal pixel scale of 110 mas pixel$^{-1}$;][]{2015BouchetMIRI}, but also the top-left region, which is designed for Lyot coronography, is also used for imaging, adding extra coverage in each exposure\footnote{See the \href{https://jwst-docs.stsci.edu/jwst-mid-infrared-instrument/miri-observing-modes/miri-imaging}{MIRI Imager JDox page} for more details.}. MIRI imaging can be performed with nine broadband filters that cover a wavelength range from 5.6 to 25.5 $\mu$m (F560W, F770W, F1000W, F1130W, F1280W, F1500W, F1800W, F2100W and F2550W; for a detailed description of MIRIM and its commissioning, see \citealt{{2015BouchetMIRI}} and Dicken et al., in preparation).

\hst and NIR \jwst's ePSFs were obtained with specific observations of relatively-crowded fields that provide thousands of stars per image for the ePSF modeling \citep[e.g.][]{2004AndersonHRCPSF,2006AndersonWFCPSF,2022NardielloNIRCam,2023LibralatoNIRISS}. At MIR wavelengths, the number of bright unsaturated sources dramatically drops, the background increases and the sensitivity drops at the longest wavelengths due to a decrease in the quantum efficiency of the detectors and significant thermal emission from the telescope. All these characteristics pushed us to try to compensate for the low statistics per exposure by including more images spanning a relatively large temporal baseline. Our MIRI ePSF models were made using publicly-available data taken during Commissioning, Cycle-1 GO and calibration programs\footnote{Program IDs: 1024, 1027, 1028, 1029, 1037, 1040, 1051, 1090, 1171, 1448, 1473, 1518, 1521, 1522, 1536, 1538, 2221.}. Only a subset of these images were used to compute the GD corrections, specifically those of the Program ID (PID) 1521. This program targeted a field in the Large Magellanic Cloud (LMC) after Commissioning (when all the major mirror alignment and phasing operations ended). This field contains a relatively-high number of sources to use for computing the GD correction. The majority of the data was taken between March and December 2022, and eight exposures from 2023 observations were included to increase the statistics in some long-wavelength filters. The number of groups, integrations and exposures differ from program to program. Overall, about the 77\% of the images were taken with $\le$20 groups and $\le$2 integrations.

We downloaded the level-1b, uncalibrated (\texttt{\_uncal}) products from the Mikulski Archive for Space Telescopes (MAST)\footnote{\href{https://mast.stsci.edu/portal/Mashup/Clients/Mast/Portal.html}{https://mast.stsci.edu/portal/Mashup/Clients/Mast/Portal.html}.}.
These exposures were processed using a development version of the \jwst pipeline\footnote{\href{https://github.com/spacetelescope/jwst}{https://github.com/spacetelescope/jwst}.} \citep{2023BushouseJWSTpipeline} through the stages 1 and 2 to obtain the level-2b (\texttt{\_cal}) fits files, fully calibrated and unresampled images \citep[see][and Dicken et al., in preparation]{2023MorrisonMIRI}. The data sets were processed at different times, so the versions of the calibration pipeline\footnote{The MIRI data was processed using either one of the following pipeline versions: 1.9.5.dev29+g40f282e, 1.9.5.dev26+g4285c4e, 1.10.2.dev11+g3f269f5.} and of some reference files\footnote{The images used in this work where processed with either one the following CRSD contexts: jwst\_1046.pmap,  jwst\_1062.pmap, jwst\_1075.pmap, jwst\_1080.pmap, jwst\_1084.pmap, jwst\_1089.pmap.} were not the same but did not impact our analyses.

\section{ePSF modeling}\label{sec:psf}

The MIRI ePSFs are Nyquist sampled or better at wavelengths $\gtrsim$ 7$\mu$m, leaving only the ePSF at 5.6 $\mu$m~with F560W to be slightly undersampled. Undersampled PSFs require careful modeling to remove the so-called pixel-phase biases (positions that are systematically measured at a specific location with respect to the pixel boundaries, regardless of where objects truly are), which can be done by combining multiple well-dithered images of the same field \citep[for a detailed description of the modeling of the undersampled ePSF of the \jwst/NIRISS detector see][]{2023LibralatoNIRISS}.

\begin{figure}[t!]
    \centering
    \includegraphics[width=\columnwidth]{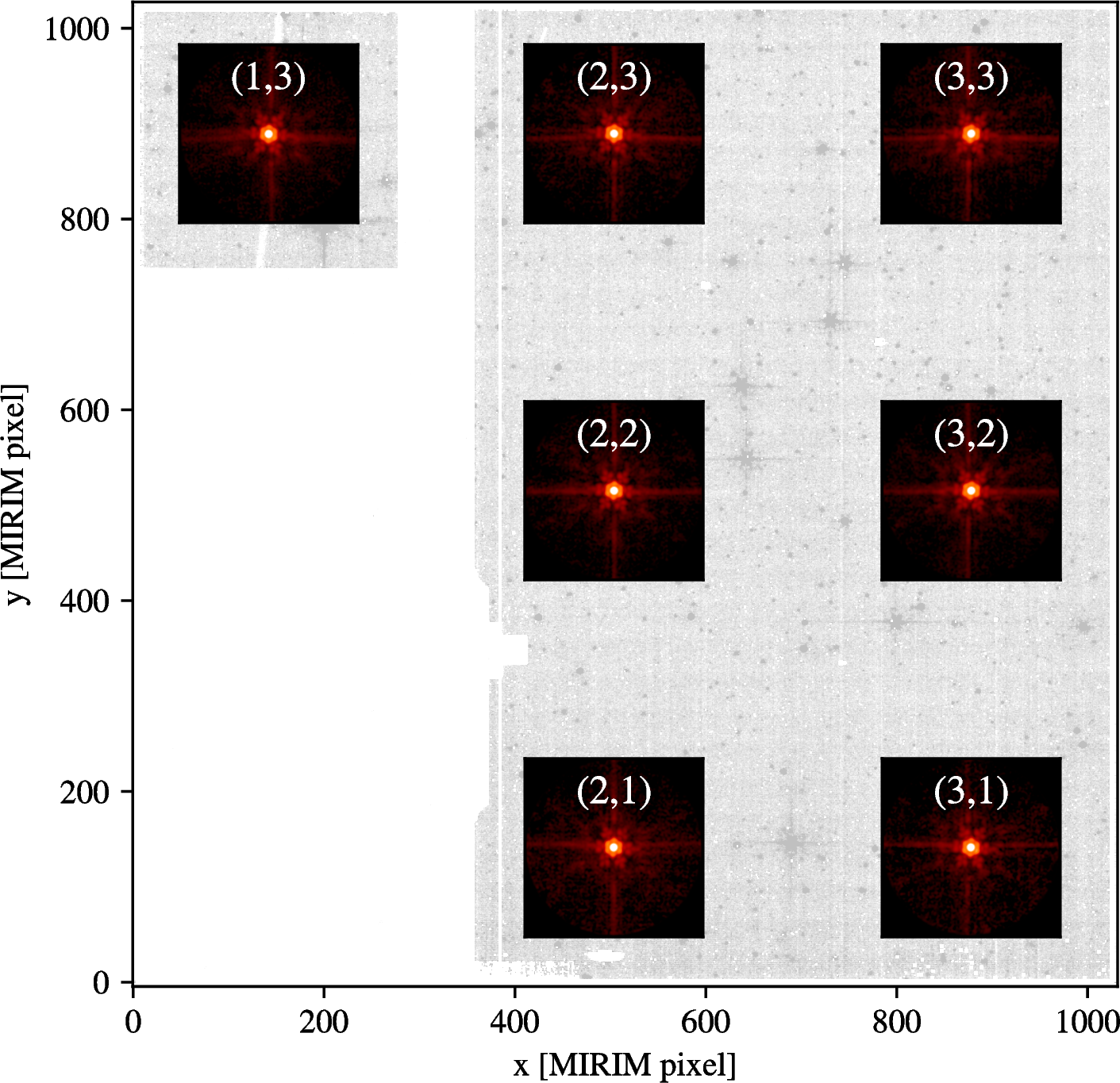}
    \caption{MIRI ePSF layout using the F560W filter data as an example. The image in the background is a \texttt{\_cal} fits file of PID 1521. The position of each ePSF is indicative of the region within which the stars for the given ePSF modeling were selected. Slots (1,1) and (1,2) are not shown because the bottom-left region of the imager is reserved for the 4-quadrant phase mask coronagraphs, thus making it not available for direct imaging.}
    \label{fig:ePSF_scheme}
\end{figure}

Because the F560W MIRI ePSF is slightly undersampled  (full width at half maximum of 1.88 pixels), we adopted a different approach from what is usually done for more severe cases: We collected the ePSF sampling from multiple stars in hundreds of images of various fields all at once and took advantage of the fact that dithers, the scene itself and even the variety of stellar fluxes of the sources in the field should distribute stars randomly with respect to the boundaries of the pixel. This solution is similar to how ePSFs are modelled with well-sampled, ground-based data \citep[e.g.,][]{2006AndersonWFI,2014LibralatoHAWKI,2015LibralatoVIRCAM,2019KerberGSAOI,2021HaberleNACO} rather than with undersampled, space-based images \citep{2000AKWFPC2}.

A pixel $(i,j)$ close to the center of a star with position $(x_{\ast},y_{\ast})$ and flux $z_{\ast}$ has a value $P_{i,j}$ that can be defined as:
\begin{displaymath}
  P_{i,j} = z_{\ast} \cdot \psi_{\rm E}(i-x_{\ast},j-y_{\ast}) + sky_{\ast}
  \phantom{1} ,
\end{displaymath}
where $sky_{\ast}$ is the local sky background (measured in an annulus region close to the source). $\psi_{\rm E}(i-x_{\ast},j-y_{\ast})$ is the fraction of the light of the stars that falls on the pixel $(i,j)$ according to the ePSF model. By inverting the equation above, we can define:
\begin{displaymath}
    \psi_{\rm E}(i-x_{\ast},j-y_{\ast}) = \frac{P_{i,j}-sky_{\ast}}{z_{\ast}} \phantom{1} .
\end{displaymath}
This simple relation tells us that every pixel in the vicinity of a star can potentially be used to sample the ePSF. Collecting many samplings from many stars in different images allows us to map the pixel-phase space and build accurate ePSF models.

For every filter, we began by measuring all bright\footnote{For the short-wavelength filters, we included stars with at least a flux of 100 DN s$^{-1}$ within the centermost $5 \times 5$ pixel$^2$. For the long-wavelength filters, we increased the threshold ($>$3000 DN s$^{-1}$) to remove spurious detections.} stars in each image. Positions $(x_{\ast}, y_{\ast})$ were initially defined as the center of light of the flux distribution in the core of each star, while fluxes $z_{\ast}$ were estimated with aperture photometry. Once ePSF models were available, positions and fluxes were computed via ePSF fit.

Our ePSF models are oversampled by a factor of 4. Each of our ePSFs is centered at a location (151,151), is normalized to have unity flux within a radius of 10 MIRI pixels (40 oversampled pixels) and extends out to 37.5 MIRI pixels (150 oversampled pixels). Smoothness and continuity of our ePSFs \citep[see][]{2000AKWFPC2} were achieved by convolving the ePSFs with low-pass smoothing kernels that removed some high-frequency structures at larger radii, while preserving the shape of real features of the ePSFs. The choice of these kernels was reached with a trial-and-error approach, and tailored for each filter (the ePSF shape changes from 5.6 to 25.5 $\mu$m).

\begin{figure*}[th!]
    \centering
    \includegraphics[width=\textwidth]{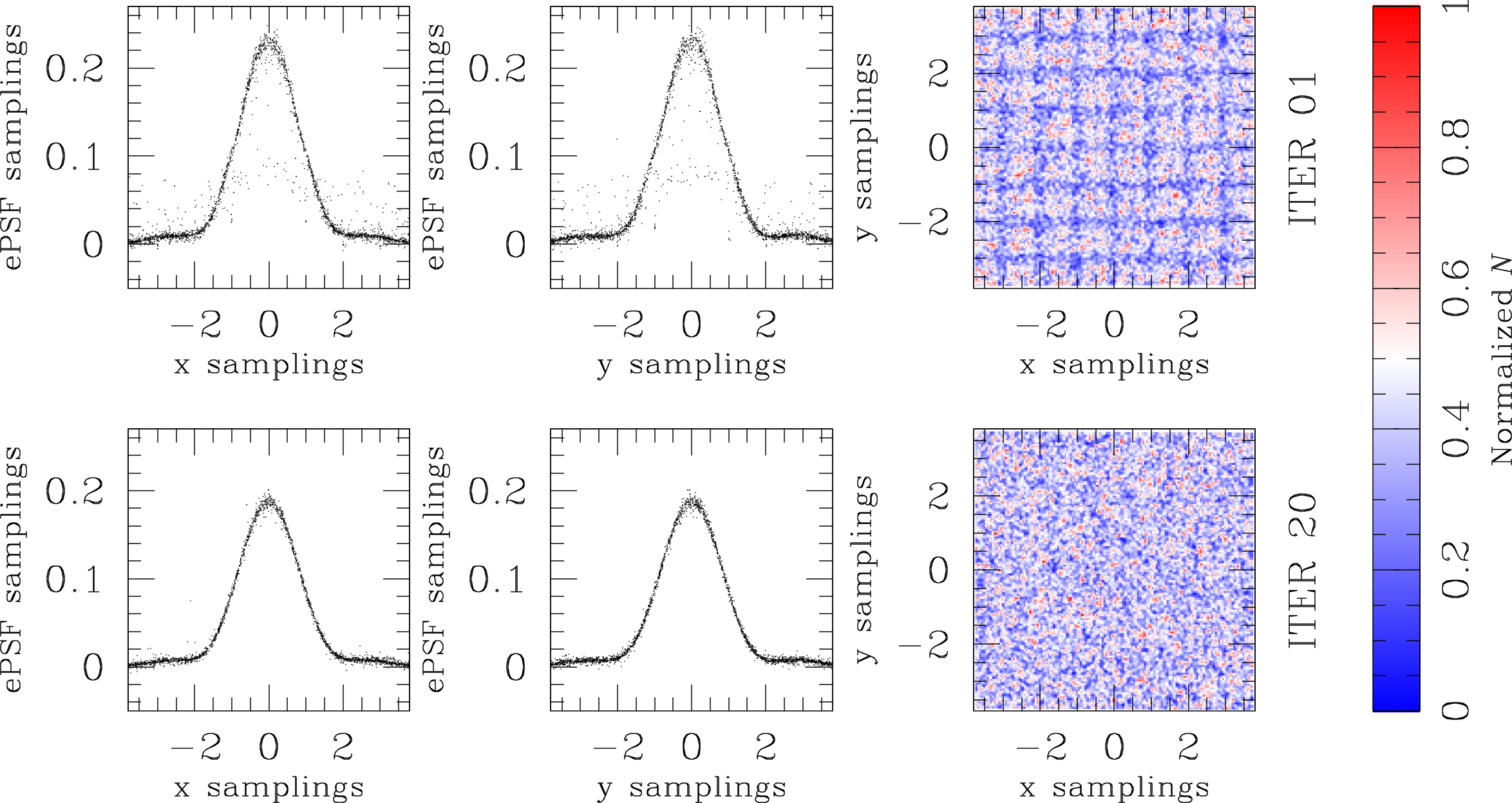}
    \caption{ePSF samplings for the F560W filter at the beginning (top row) and at the end (bottom row) of our iterative ePSF-modeling process. The first two columns from the left show the ePSF samplings ($\psi_{\rm E}$) as a function of distance from the center of the ePSF along the $x$ and $y$ axes, respectively. Only 1\% of the points in a strip within $\pm$0.1 pixel from the center of the stars are plotted. The density plots in the third column (color-coded as in the colorbar on the right) highlight where the measured centers of the stars are with respect to the boundaries of the pixels. At the beginning of the iterative process, the ePSF samplings are preferentially measured at the corners of the pixels, and the ePSF model is too sharp and with many outliers. Once accurate ePSF models are used to find the center of the stars, the ePSF samplings are more homogeneously distributed with respect to the pixel boundaries, and the shape of the ePSF is smooth and well constrained.}
    \label{fig:f560w_sampling}
\end{figure*}

We started with a single ePSF model for the entire detector, and progressively let the ePSFs to vary spatially. We found that one ePSF that covers the Lyot region and a 2$\times$3 array of ePSFs (Fig.~\ref{fig:ePSF_scheme}) for the imager were a good compromise between having enough stars in each region of the detector to model the ePSF and including spatial variations of the ePSF across the FoV. This choice allows us to interpolate the ePSF at any given pixel of the imager region, but forces fitting the same ePSF in all pixels in the Lyot region. Because the number of stars drastically drops from short to long wavelengths, the spatial variability cannot be included in all filters. Also, sometimes this same issue did not let us model the spatial variability in the wings of the ePSFs, and imposed ePSF models with a spatially-variable core and constant wings. The details of the spatial variability of our ePSF are summarized as follows:
\begin{itemize}
    \item F560W and F770W: one ePSF for the Lyot and 2$\times$3 ePSFs for the imager. These ePSF models are spatially variable both in the core and in the wings, so to include the change of the location and intensity of the so-called cruciform artifact \citep[e.g.,][]{2021PASP..133a4504G};
    \item F1000W, F1130W\footnote{Most of the stars used to model the wings of the F1130W ePSFs are embedded in dust, which caused a gradient in the wings of the ePSFs in this filter. We advise caution when using the wings of these particular ePSFs.} and F1280W: one ePSF for the Lyot and 2$\times$3 ePSFs for the imager. These ePSF models are spatially variable in the core out to 7.5 pixels (30 oversampled pixels), but constant in the wings (i.e., all ePSFs share the same profile in the wings);
    \item F1500W: one ePSF for the Lyot and 2$\times$2 ePSFs for the imager (the remaining two ePSFs were linearly interpolated by means of the ePSFs at the corners of the imager). As for the previous filters, these ePSF models are spatially variable in the core out to 7.5 pixels, but constant in the wings;
    \item F1800W, F2100W and F2550W: one ePSF model for the entire MIRI detector (Lyot and imager).
\end{itemize}
The modeling of the MIRI ePSF was made with an iterative process. At each iteration, position and flux of all sources were measured with the available ePSFs, then the ePSF samplings were redefined and the ePSFs updated. Convergence was reached between 10 (one ePSF and no spatial variability) and 20 (all ePSFs and spatial variability) iterations.

\begin{figure}[t!]
    \centering
    \includegraphics[width=\columnwidth]{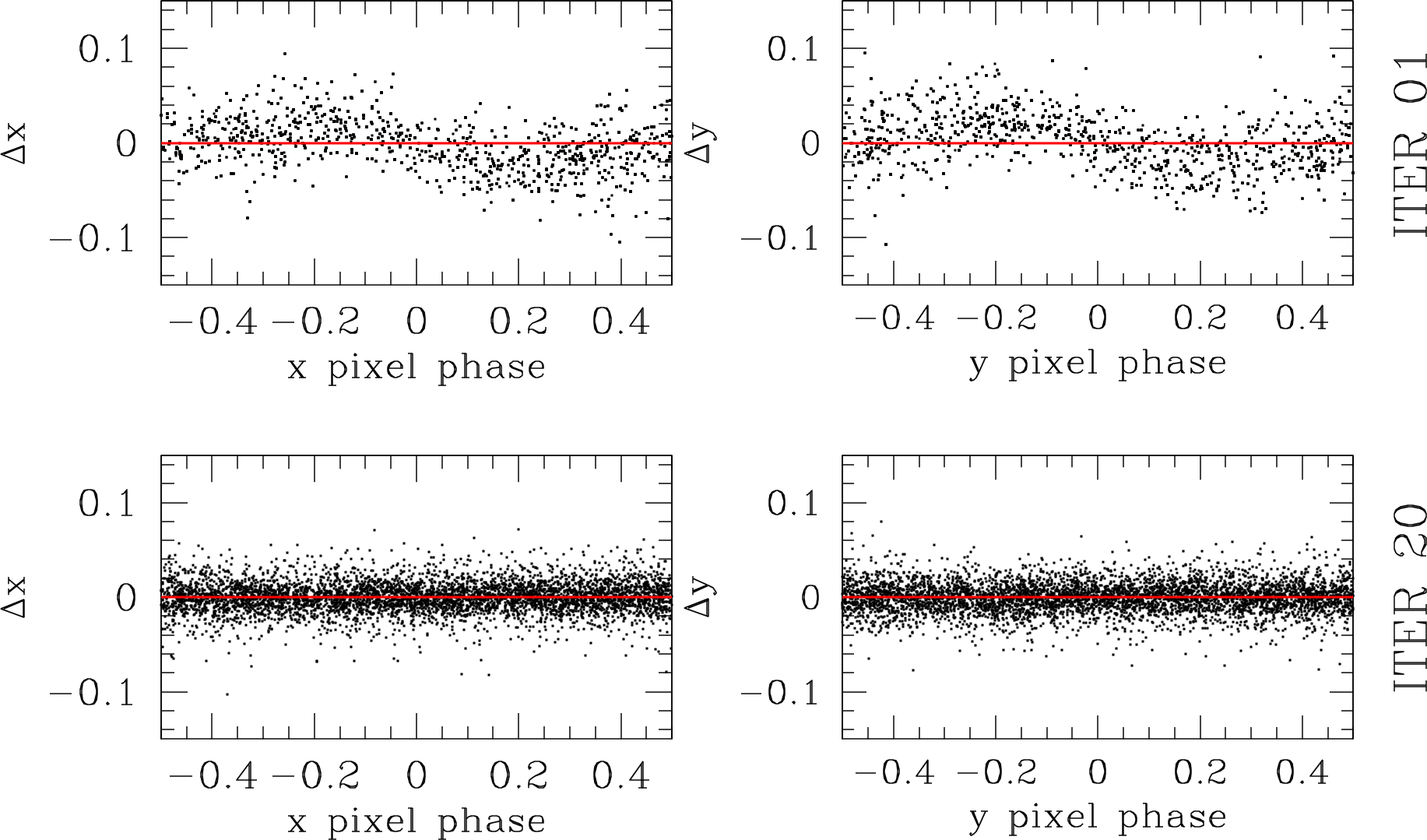}
    \caption{Pixel-phase errors for the $\Delta x$ and $\Delta y$ positional residuals. The top row refers to the first iteration of our ePSF modeling when the center of the stars was defined as the light-center of the flux distribution. The bottom row shows the results for the last iteration when the ePSF models were fit to compute the position of the stars. The same stars are shown in all panels.}
    \label{fig:f560w_pp}
\end{figure}

Figure~\ref{fig:f560w_sampling} shows how the ePSF samplings change for the F560W case, the only undersampled ePSF in the MIRI imager. At the first iteration, where positions were defined as the light-center of the flux distribution and fluxes were computed via aperture photometry, the shape of the ePSF was not well constrained, as stars were preferentially measured to be at the corners of the pixels (top panels). Thanks to our careful modeling, the shape of the ePSF becomes smoother and with less scatter than before. Also, stars do not show any preferred positioning in the pixel-phase space (bottom panels).

\begin{figure*}[th!]
    \centering
    \includegraphics[width=\textwidth]{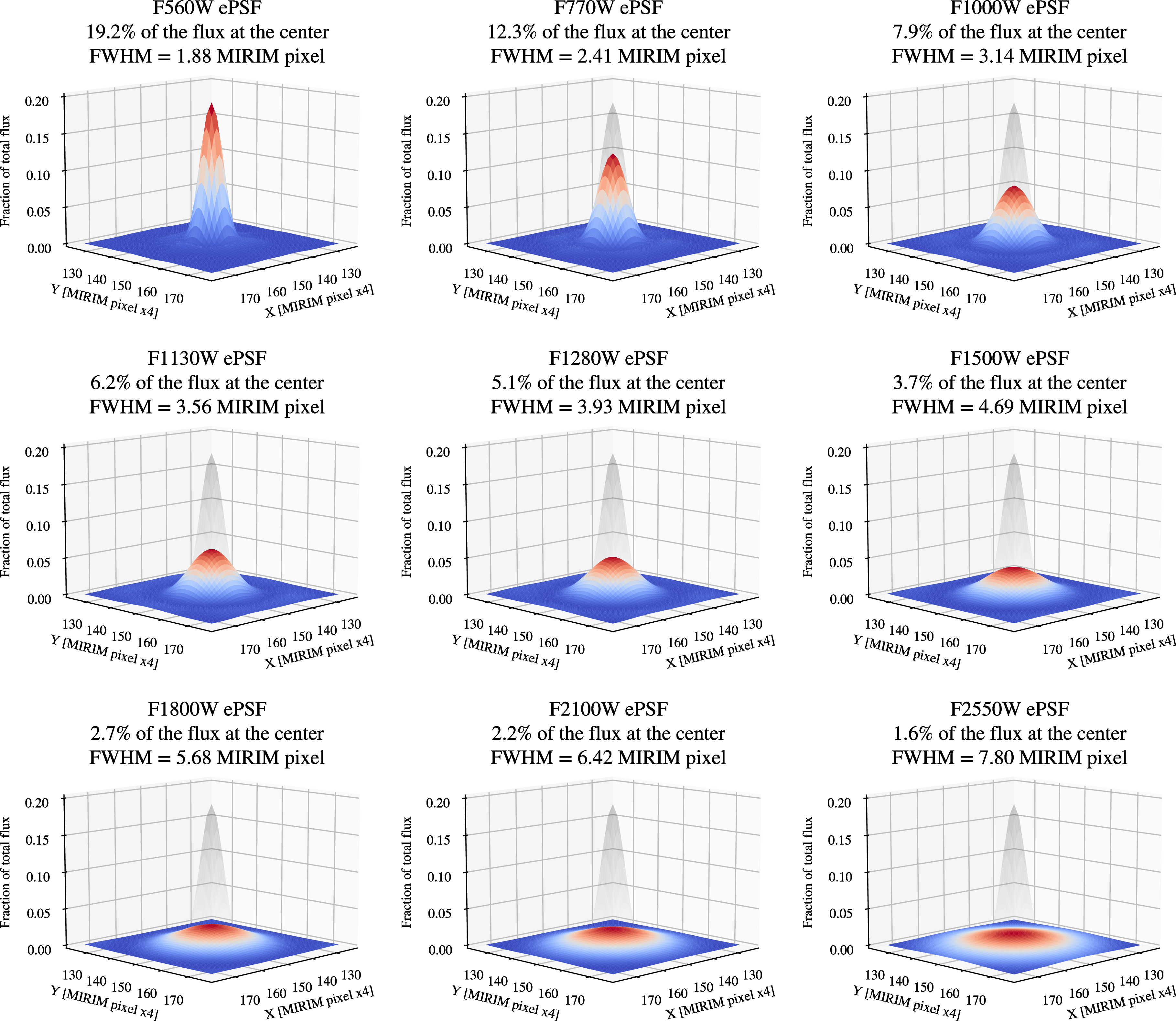}
    \caption{3D view of our ePSF models. In each panel, the colored surface represent the ePSF in the filter reported on top, while the gray surface is the profile of the F560W ePSF as a reference. At the top of each panel, we provide the percentage of flux of the star within the centermost pixel and the full width at half maximum (FWHM) in pixel.}
    \label{fig:ePSF_3D}
\end{figure*}

Another way to assess the impact of the pixel-phase errors introduced by non-optimal ePSFs can be obtained by using dithered observations \citep{2000AKWFPC2}, because dithering places a source at different locations with respect to the pixel boundaries. By combining positions of stars measured in multiple dithered images, averaging them together on to the same reference-frame system, we can obtain a set of positions free from pixel-phase systematic errors, and these positions can be compared with those measured in each MIRI raw frame \citep[e.g.,][]{2023LibralatoNIRISS}. This test was run for the F560W filter.

We measured positions and fluxes of stars in the LMC calibration field and other adjacent fields observed during Commissioning and Cycle-1 Calibration programs. We cross-identified the same stars in multiple catalogs\footnote{In the cross-matching process, we took advantage of GD corrections described in Sect.~\ref{sec:gd}, so that pixel-phase errors are not hidden by the systematics related to the GD, which are $\sim$100 times larger.}, and averaged their positions and fluxes after they were transformed on to a common reference frame system, hereafter simply master frame, by means of six-parameter linear transformations. The scale and orientation of the master frame were set up by means of the \gaia Early Data Release 3 (EDR3) catalog \citep{2016GaiaCit,2021GaiaEDR3} projected on to a tangent plane centered at (R.A.,Dec.) $=$ (80.606608, $-$69.461994) deg, fixing the pixel scale to be the nominal pixel scale of MIRI (110 mas pixel$^{-1}$). The photometry of our master frame was registered to that of a MIRI catalog.

Finally, we compared the raw positions of bright, well-measured stars in each MIRI image with those predicted by the averaged master frame inverted on to the MIRI raw frame. Figure~\ref{fig:f560w_pp} illustrates the impact of pixel-phase errors in astrometry. In the top panels, the positional residuals show a clear trend as a function of pixel phase when positions were defined as the light-center of the flux distribution. Pixel-phase biases disappear when stellar positions are fit using our ePSF (bottom panels).

Figure~\ref{fig:ePSF_3D} presents a 3D overview of one ePSF for each filter. The colored profiles refer to the ePSF of each filter, whereas the gray surfaces are the 3D profile of the F560W ePSF for reference. The ePSFs becomes progressively less sharp (from the 19.2\% of the total flux in the centermost pixel of the F560W ePSF to the 1.6\% of the F2550W ePSF) and broader (from the full-width half maximum of 1.88 pixels of the F560W ePSF to the 7.80 pixels of the F2550W ePSF) from short to long wavelengths. 

The \texttt{python} package WebbPSF \citep{2012PerrinWebbPSF,2014PerrinWebbPSF} can provide simulated PSFs that in principle can be used for astrometry and photometry with the \jwst's imagers. However, at the moment WebbPSF does not include all detector features that contribute to the effective PSF. Appendix~\ref{sec:webbpsf} describes a comparison between our ePSFs and WebbPSF PSFs. Our result shows that caution is advised when using WebbPSF models, at least for now, especially for the F560W and F770W filters.

\subsection{Quality of the ePSF fit}\label{sec:psfcheck}

To assess the efficacy of our ePSFs, we made use of two parameters: the ``Quality of PSF fit'', or \qfit, and the ``radial-excess'' parameter, or \radxs.  The \qfit is defined as the absolute fractional error in the ePSF fit of a source \citep[e.g.,][]{2006AndersonWFI,2014LibralatoHAWKI}. The closer the \qfit is to zero, the better is the fit of the ePSF. The \radxs is defined as the the excess/deficiency of flux outside the core (between 1 and 2.5 pixels) of the source compared to the prediction of the ePSF model \citep{2008ApJ...678.1279B}. Stars usually have a \radxs close to 0; cosmic rays or hot pixels have a negative \radxs value (i.e., they appear sharper than the ePSF model); and galaxies exhibit positive \radxs values (they are broader than the ePSF model). Thus, the \radxs is usually a powerful parameter to understand what kind of sources we are dealing with.

\begin{figure*}[th!]
    \centering
    \includegraphics[height=15cm, keepaspectratio]{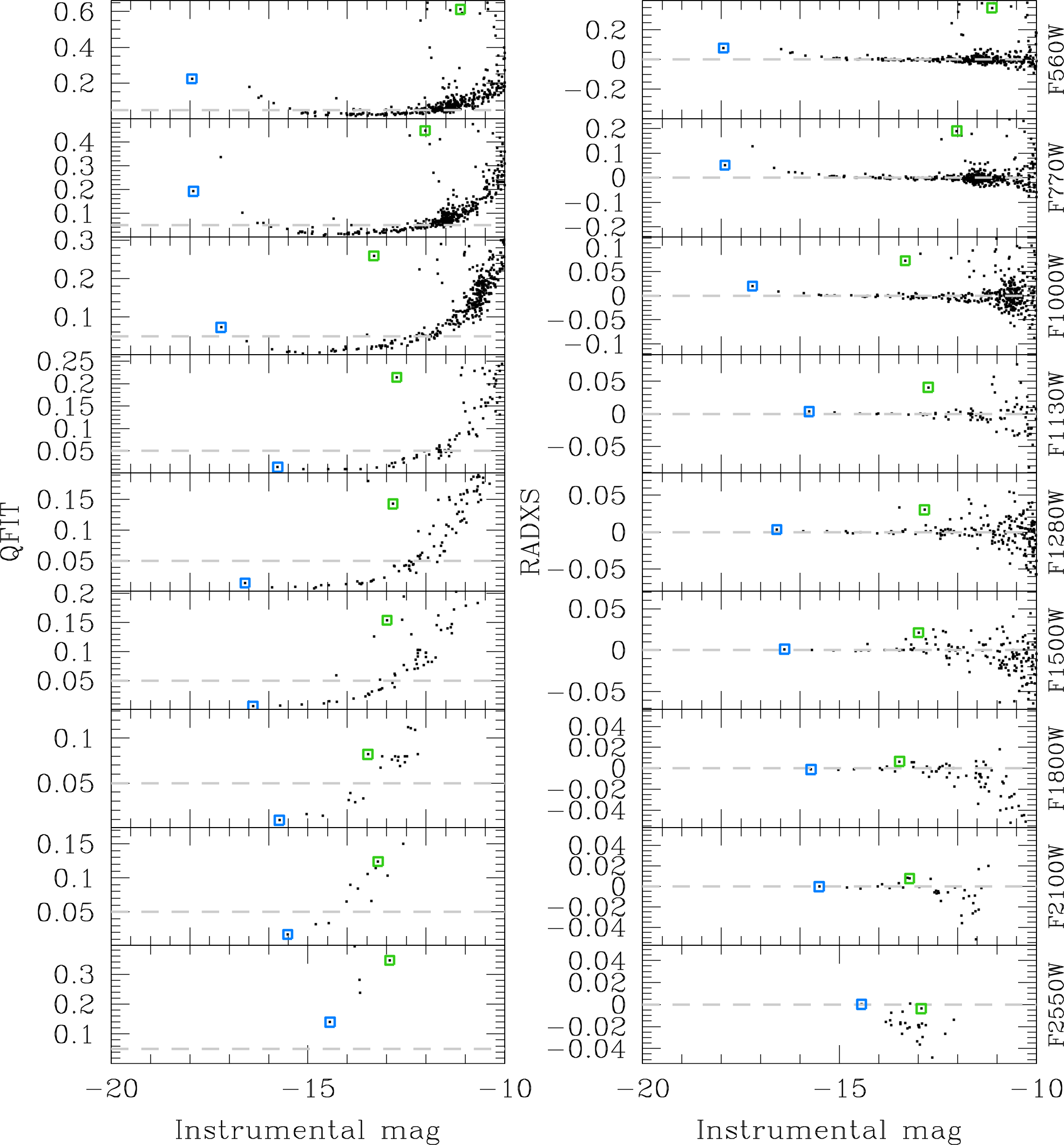}~\hskip 15pt
    \raisebox{1.cm}{\includegraphics[height=14cm, keepaspectratio]{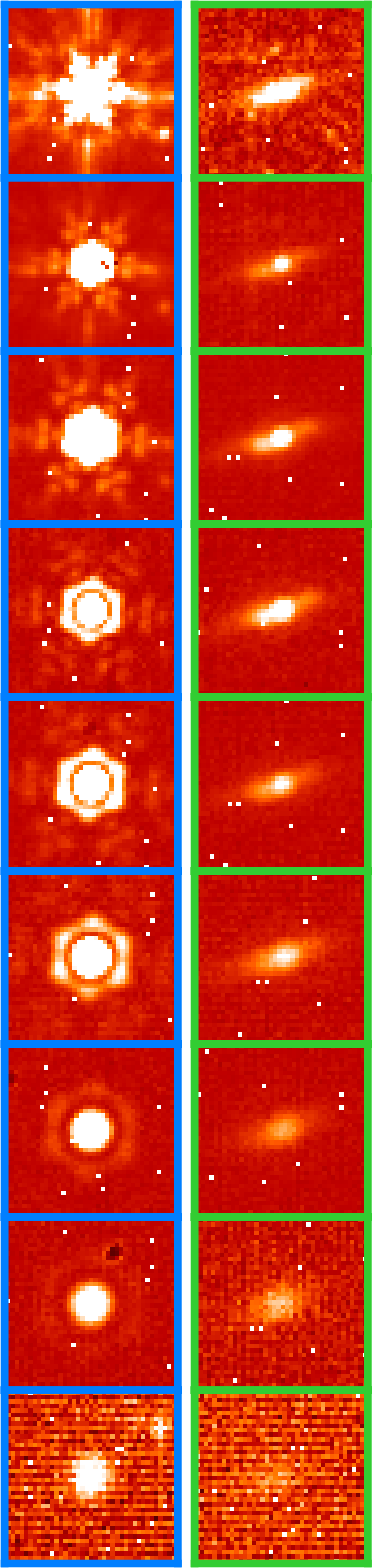}}
    \caption{Quality of the ePSF fit in each filter for a representative image of PID 1024. The left panels show the \qfit as a function of instrumental magnitude. The horizontal, dashed, gray line is set at \qfit $=$ 0.05. The middle panels provide analogous trends but for the \radxs. The  horizontal, dashed, gray line is set at \radxs$=$ 0. The blue and green squares highlight the position of a representative star and galaxy in each plot, respectively. The rightmost panels show these two objects in each image. Structures in the lowermost right panels are due to the high noise in the image. White pixels are those flagged by the pipeline as not to use.}
    \label{fig:qfit_radxs}
\end{figure*}

We run this test on all images, but discuss the result for the Commissioning program PID 1024, which observed the LMC calibration field in all filters. Figure~\ref{fig:qfit_radxs} shows \qfit and \radxs as a function of instrumental magnitude for one image in each filter. The instrumental magnitude is defined as $-2.5 \log (\textrm{DNs})$, where DNs are the total accumulated Digital Numbers in the exposure. These values were obtained by converting the pixel values in MJy sr$^{-1}$ (the physical unit of the level-2b files) in DN s$^{-1}$ using the conversion factor provided by the header keyword \texttt{PHOTMJSR}, and then multiplying the result by the effective exposure time available in the header keyword \texttt{EFFEXPTM}. Although the actual exposure time of each pixel can differ depending on the actual groups used in the ramp fit, the DN values here computed are a reasonable proxy of the signal of our sources.

The left panels highlight the typical trend of the \qfit as a function of instrumental magnitude. The horizontal, gray, dashed line is set at \qfit $=$ 0.05, a value generally used to select well-measured stars. The middle panels present analogous plots but for the \radxs. The horizontal line is instead set at \radxs $=$ 0, the expected value for a point-like source. With the exception of the F2550W filter, for which this specific data set does not contain any bright star, we can measure bright point-like sources well (low \qfit and \radxs $\sim$ 0) thanks to our ePSF models. At faint magnitudes, the \qfit increases and the \radxs distribution becomes broader because the signal-to-noise ratio of the sources decreases, and so objects becomes progressively poorly measured, but the points are mostly centered at \radxs $\sim$ 0. Very-bright stars in the short-wavelength filters show an increasing \qfit and a positive trend for the \radxs. This behavior is likely due to a combination of various factors, including saturation, non-linearity \citep[although it is expected to be small;][]{2023MorrisonMIRI} and, most importantly, brighter-fatter \citep[there is more flux just outside the core of the star, meaning the object is ``fatter'' than what the ePSF predicts;][see also Sect.~\ref{sec:psfbf}]{2023arXiv230313517A}. Similar trends have been found for the NIRISS detector \citep{2023LibralatoNIRISS}.

The sensitivity and wavelength range covered by the MIRI imager make it possible to easily find galaxies in almost every exposure. To further help readers to understand the result of ePSF fit with MIRI data, and how it differs for point-like and extended sources, we selected a star and a galaxy which were observed in all our images in this test. The choice fell on two specific objects, i.e., the brightest sources detectable in the F2550W images. In all panels in Fig.~\ref{fig:qfit_radxs}, the star is highlighted with a blue square, while the galaxy is depicted with a green square. The rightmost two panels, using this same color code, are zoomed in around these sources in each image.

For wavelengths $\leq$10 $\mu$m, the star is very bright and its \qfit and \radxs values are larger than what we would expect for a well-measured star for the reasons mentioned above. The galaxy has instead a large \qfit and a positive \radxs value, which place it in a different location with respect to where stars are.

From 11.3 to 21 $\mu$m, the star is now well below the line set at \qfit$=$0.05, meaning that our ePSF models are adequate. On the other hand, the galaxy progressively displays \qfit and \radxs values closer to those of the stars. At 25 $\mu$m, the situation becomes even more extreme, with star and galaxy sharing similar \radxs values. This is likely due the larger FWHMs of the ePSFs at these wavelengths. Specifically, the longer the wavelength, the shallower and broader the ePSF becomes, and so differences in the \qfit and \radxs of point and extended sources are more subtle. For this reason, we advise to visually inspect the targets in the image as an additional cross-check.

\begin{figure*}[th!]
    \centering
    \includegraphics[width=\textwidth]{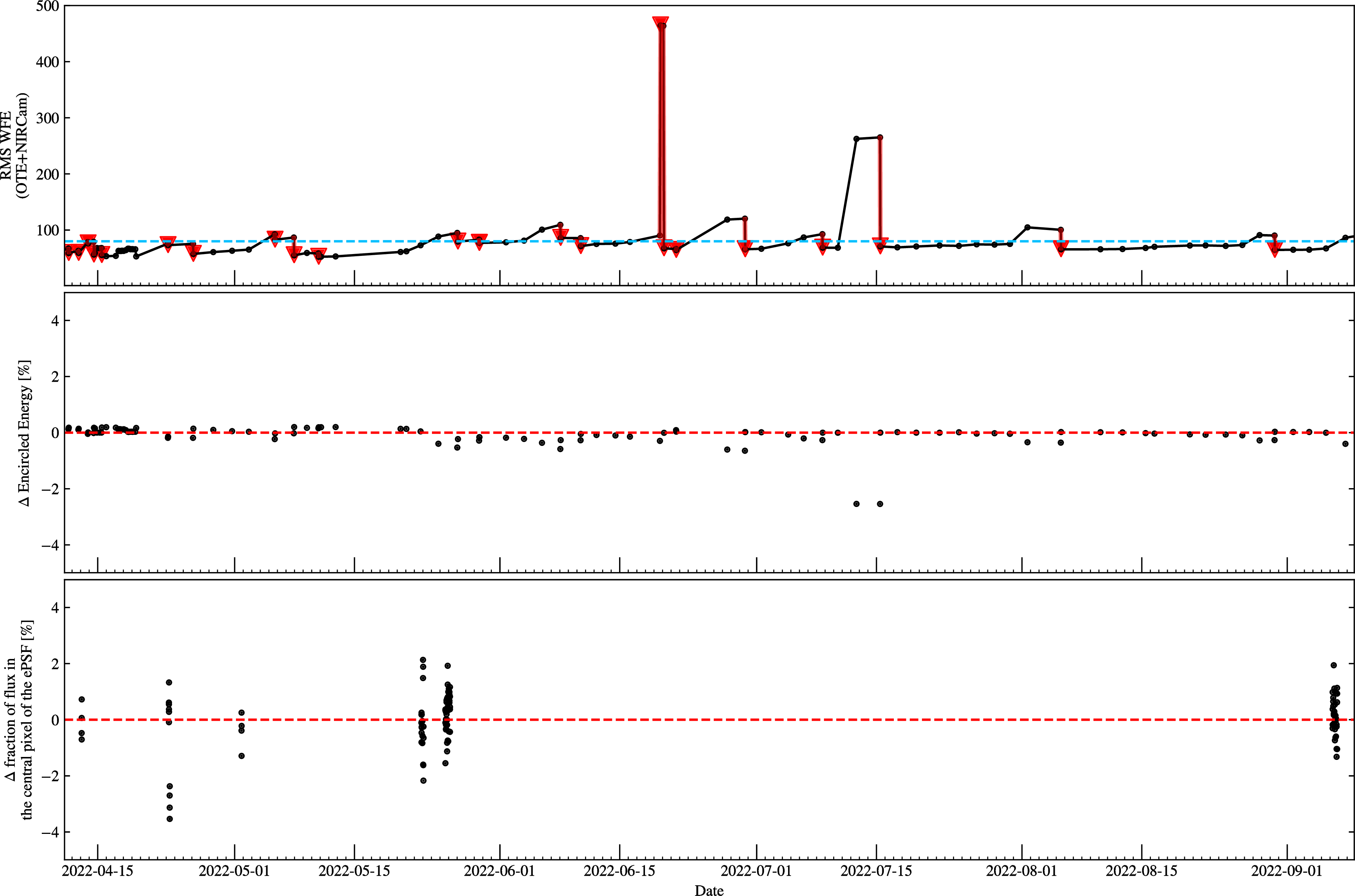}
    \caption{In the top panel, we show the rms of the wave front error (WFE) of the observatory as a function of time \citep[see also][]{2023McElwainJWST}. The black dots mark the mirror sensing visits. The red arrows highlights when the primary mirror segments assemblies were corrected, because the rms WFE was above the correction threshold (light-blue, dashed, horizontal line). The variations in the Encircled Energy (EE) of the F560W MIRI WebbPSF PSFs retrieved from the OPD files is presented in the middle panel. The red, dashed, horizontal line is set at 0 as a reference. Finally, the change of the fraction of light that falls in the centermost pixel of our F560W ePSF at the center of the FoV is plotted in bottom panel (again, a reference red, dashed, horizontal line is set at 0).}
    \label{fig:f560w_ote}
\end{figure*}

\subsection{Temporal stability of the ePSFs}\label{sec:psftime}

In the following, we refer to the final ePSF models described in the previous Sections as the library ePSFs. Our work made use of images taken over a temporal baseline of a few months. Although limited, this coverage allows us to monitor if, and by how much, the ePSFs change over time. The same limitations discussed for the library-ePSF modeling applies here: the low statistics makes this monitoring feasible only for the short-wavelength filters (F560W, F770W, F1000W) and using images that cover fields in the LMC, where the stellar density is moderate.

We fined-tuned our library ePSF models for every image as described in, e.g., \citet{2017BelliniwCenI} or \citet{2018LibralatoNGC362}: We iteratively ePSF-fitted and subtracted well-measured, bright stars in the image, and adjusted the ePSF models to minimize the residuals of the ePSF subtraction. Again, the number of stars at our disposal for this task in each image is low, so we collected the residuals over the entire FoV of MIRI and perturbed all ePSFs in the array by the same quantity (so potential spatial dependencies of the temporal variations are not considered). We find that the ePSF variations are minimal, below 4\% in the worst case. These small variations are in agreement with what was found over a similar temporal baseline for NIRCam \citep{2022NardielloNIRCam} and NIRISS \citep{2023LibralatoNIRISS} imagers.

We made use of WebbPSF to correlate the variations we measured in the centermost pixel of the ePSFs with the in-flight wavefront sensing measurements. We retrieved the Optical Path Difference (OPD) files, which include information about the mirrors for the period covered by the observations used to monitor the change of the ePSFs. We looked at the variations over time of the telescope$+$instrument wavefront rms value and of the encircled energy (EE) of MIRI, and compared them with the change in the peak of our ePSFs (we chose the centermost ePSF as a benchmark). The result, presented in Fig.~\ref{fig:f560w_ote}, shows that the mirror configuration did not change significantly in our observations, thus confirming the lack of temporal variations of the ePSFs we measured.

As described in \citet{2023LibralatoNIRISS}, the perturbation process mainly affects the photometry, while astrometry is not expected to improve significantly
\footnote{Some changes in the reference files and pipeline used in the processing of the data can also require a perturbation of the library ePSFs. However, turning on/off some pipeline steps like the \href{https://jwst-pipeline.readthedocs.io/en/latest/jwst/ipc/description.html}{inter-pixel-capacitance correction} can significantly impact how the flux of point sources is distributed across pixels and could require new ePSF models.}. Our tests show that the ePSFs are stable, as expected given the overall stability of \jwst at these wavelengths. Users should always carefully explore whether the perturbation procedure actually improves the ePSFs, and should use many high signal-to-noise stars to fine-tune the ePSFs.

\subsection{A note on the brighter-fatter effect}\label{sec:psfbf}

\citet{2023arXiv230313517A} have firstly reported evidence of Brighter-Fatter-like (BF) effects in MIRI. When a star is impacted by this effect in the MIRI imager, its profile becomes broader and less peaked than normal (see their Fig.~6). We investigated  whether, and to what extent, the BF effect impacts our ePSF models and the photometry resulting from their fit.

We followed a similar approach to that of \citet{2023arXiv230313517A}. Briefly, we used F560W data of the Commissioning program PID 1464. Each of the 12 images consists of a single integration and 125 groups. Such long ramps are perfect to assess the presence of the BF effect. We processed the data three times modifying the \jwst pipeline so to use only the first 10 (i.e., the ramp of very bright stars is still not impacted by the BF), 50 (ramp mildly impacted by the BF) and all 125 (non-linear deviations towards the end of the ramp due to the BF effect) groups, respectively. We did so by changing the Data Quality (DQ) group flag of the remaining groups to ``saturated'' when running the stage-1 pipeline. This strategy allowed us to keep everything except the ramp fit as similar as possible between the three data sets. Finally, we fit our F560W ePSF models to all stars in each sample. Our ePSFs were computed with images taken with various combinations of groups and integrations, but most ramps were relatively short (Sect.~\ref{sec:data}). Also, we restricted the sample of stars for the modeling of the core of the ePSFs to objects not too bright, so to avoid dealing with non-linearity and BF effects. This means that our ePSF models should be a good representation of a point-like source not impacted by these issues.

\begin{figure}[t!]
    \centering
    \includegraphics[width=\columnwidth]{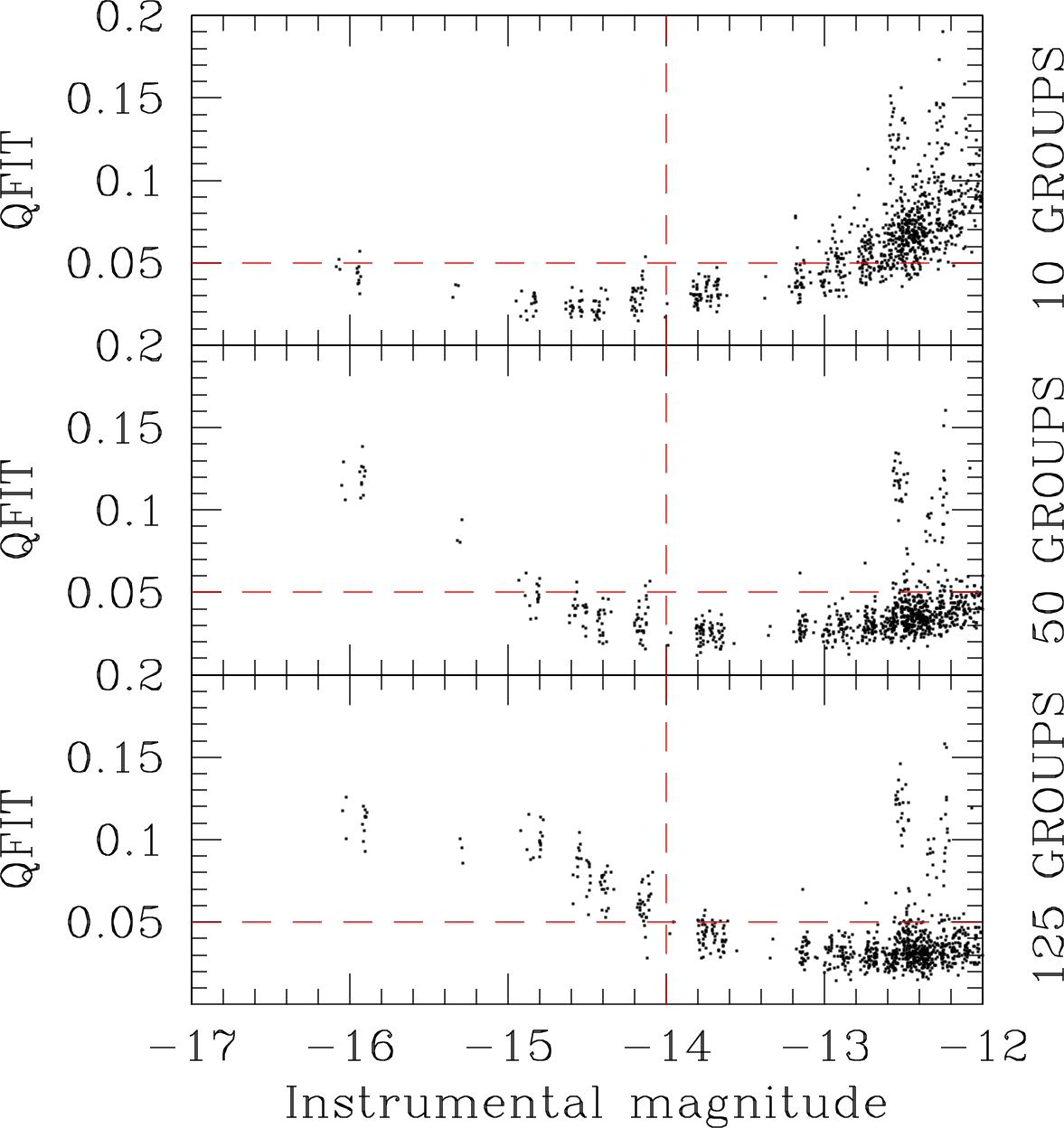}
    \caption{\texttt{QFIT} as a function of instrumental magnitude for all PID-1464 images processed using the first 10 groups (top panel), 50 groups (middle panel) and all 125 groups (bottom panel). The red, horizontal, dashed line is set at 0.05, while the vertical one is set at instrumental magnitude $-14$. These two lines are set for reference.}
    \label{fig:qfit_bf}
\end{figure}

In the first test, we looked at the \qfit parameters in the three cases. Figure~\ref{fig:qfit_bf} shows the \qfit as a function of magnitude for all images processed using 10 (top panel), 50 (middle panel) and 125 (bottom panel) groups. For simplicity, the instrumental magnitudes were obtained by assuming the same total exposure time in all cases. The main difference between the three plots happens for stars with magnitude brighter than $-14$: the higher is the number of groups used in the ramp fit, the larger is the \qfit value. This means that the profile of these bright stars progressively deviates from what predicted by our ePSFs. Stars with magnitude $>$$-13$ show a different trend when only 10 groups are fit. This is likely due the faintness of the targets (the ramp fit improves when more photons are collected and more groups are fit).

\begin{figure}[t!]
    \centering
    \includegraphics[width=\columnwidth]{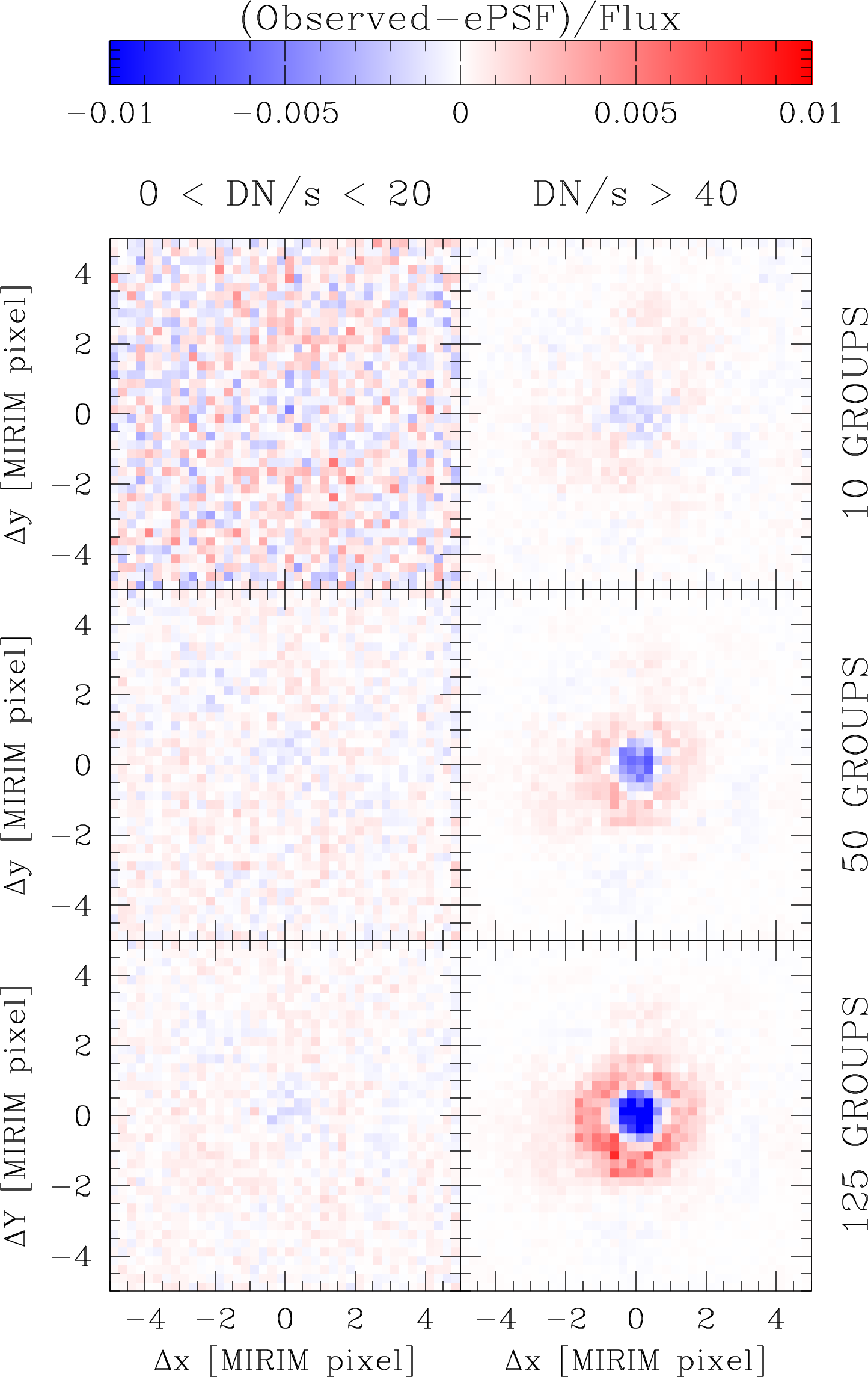}
    \caption{Overview of the BF effect in the fit of our ePSF. The left panels shows the residuals from the ePSF subtraction in the centermost 5$\times$5 pixel$^2$ for all stars in the faint regime, whereas the right panels focus on very bright sources that are likely impacted by the BF effect. From top to bottom, each row corresponds to the result obtained using images processed using only the first 10, 50 or all 125 groups in the ramp fit, respectively.}
    \label{fig:f560w_bf}
\end{figure}

The second test evaluated the residuals from the ePSF subtraction of all stars, i.e., the difference between the pixel value in the image (sky subtracted) and the value predicted by the ePSF fit in the same pixel. We collected the centermost 5$\times$5 pixels$^2$ of each star where the BF effect is expected to show up. Figure~\ref{fig:f560w_bf} presents the residuals for the three cases. Left panels refer to relatively faint stars ($<$20 DN s$^{-1}$). The residuals look homogeneously distributed in the pixel space in all three cases, which is what we expect for stars in a brightness regime not affected by the BF. Right panels show the residuals for very bright stars ($>$40 DN s$^{-1}$). We can clearly see that at the center of the star the ePSF predicts more flux than what is observed; conversely, the ePSF predicts less flux than what is present in the pixels just outside the core. The discrepancies between observed and predicted values increases the higher the number of groups used in the ramp fit. This means that these bright sources have a profile less peaked and broader than the ePSF models, as expected because of the presence of the BF effect.

\begin{figure}[t!]
    \centering
    \includegraphics[width=\columnwidth]{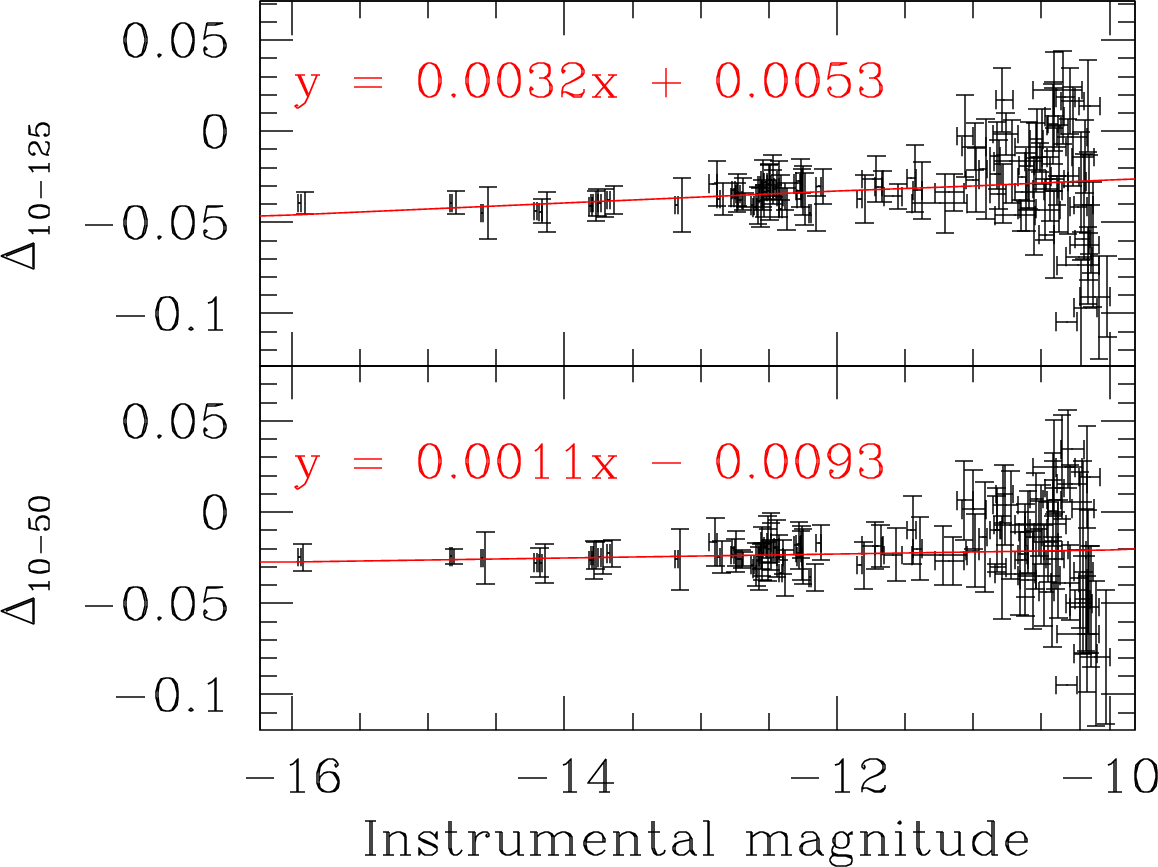}
    \caption{Impact of the BF effect on PSF photometry. The top panel shows the magnitude difference between the instrumental magnitude measured from the master frame made with images with 10 and 125 groups used in the fit, while the bottom panel refers to the comparison between the 10- and 50-group cases. Error bars are defined as the sum in quadrature of the magnitude rms of the stars in each sample. The red lines are a weighted least-square straight-line fit to the point (the equation is reported in red).}
    \label{fig:master_bf}
\end{figure}

The last test was a comparison between the photometry obtained from each set of images. For each sample, we made a master frame (similarly to what described in Sect.~\ref{sec:psf}). We kept only stars present in at least 10 images. We then cross-identified the same stars between the three master frames, and compared the magnitudes of the stars (Fig.~\ref{fig:master_bf}). There is a mild linear relation as a function of magnitude suggesting that the brighter is the star, the fainter it appears in the 50- or 125-group cases with respect to the 10-group case. Although other effects, like e.g. non-linearity\footnote{The uncertainty on the linearization of the ramps is larger in the presence of the BF effect, resulting in a larger uncertainty in the flux measured by each pixel}, could contribute to the observed discrepancies, this final piece of evidence seems to point again at the BF effect as possible cause. A correction of the BF effects to apply at the ramp level will be discussed in an ongoing work (Gasman et al., in preparation).

\section{Geometric-distortion correction}\label{sec:gd}

To solve for the GD of the MIRI imager, we made use of data from program 1521. As done for other instruments \citep[e.g.,][]{2014LibralatoHAWKI,2009BelliniWFC3,2011BelliniWFC3,2021HaberleNACO,2023GriggioNIRCam,2023LibralatoNIRISS}, our GD correction is made up by two parts: a polynomial solution and a look-up table of residuals. We computed the GD corrections for the F560W, F770W, and F1000W filters. We postpone the analysis of the other filters to future releases.

The GD corrections for the MIRI imager were computed using the \gaia EDR3 catalog as a reference. The \gaia positions were propagated at epoch mid 2022 (a representative date for our observations) to remove the contribution of the internal motions of the LMC stars in the GD computation. Then, we projected positions on to a tangent plane centered at (R.A.,Dec.) $=$ (80.606608, $-$69.461994) deg. The pixel scale was set to be the nominal pixel scale of MIRI (110 mas pixel$^{-1}$), and the $x$ and $y$ axes were oriented West and North, respectively.

\begin{figure*}[th!]
    \centering
    \includegraphics[width=\columnwidth]{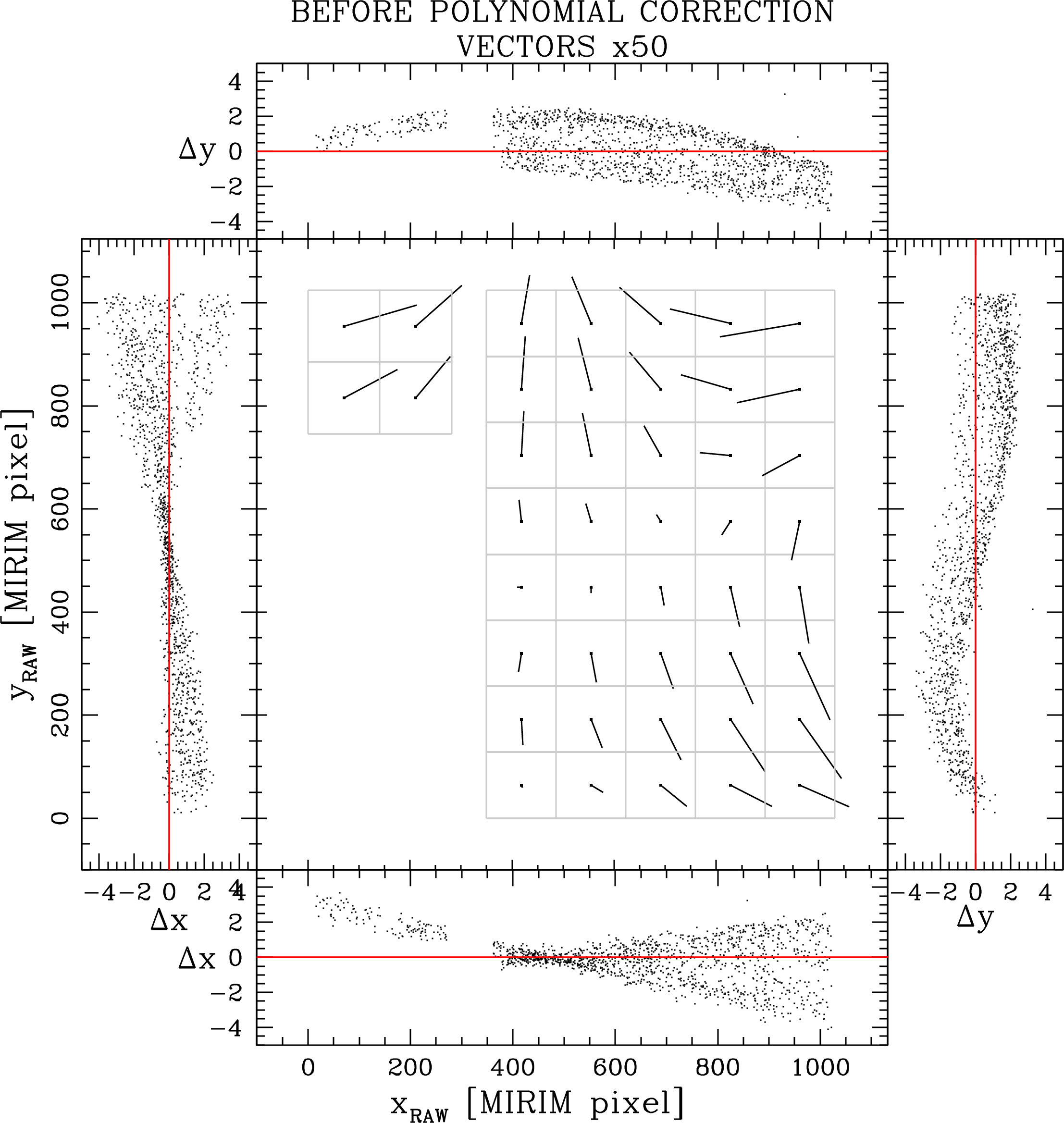}
    \includegraphics[width=\columnwidth]{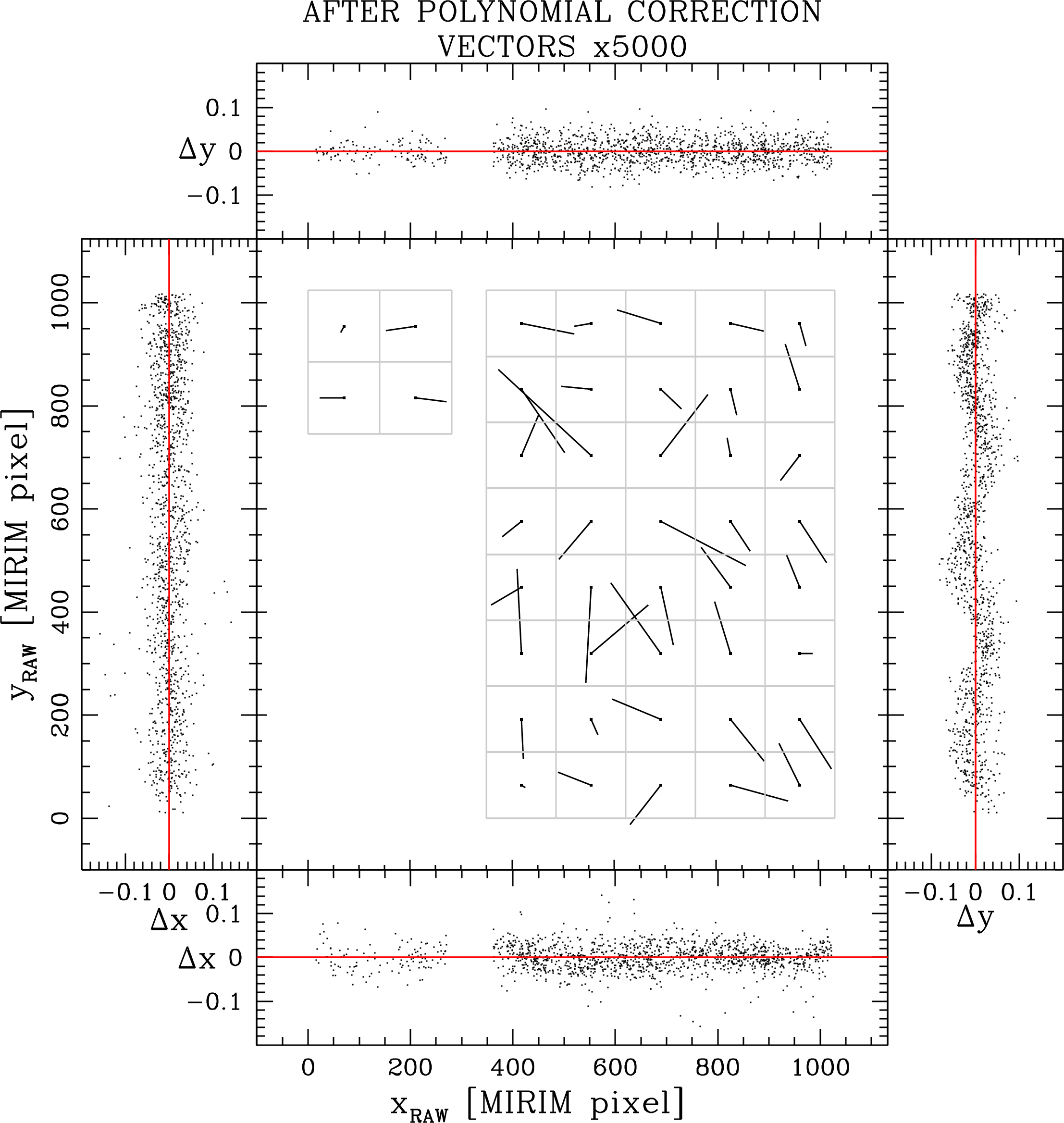}
    \caption{GD maps before (left panel) and after (right panel) applying the polynomial correction for the F560W data. Vectors are magnified by a factor of 50 (left) and 5\,000 (right) to enhance the details. The positional $x$ and $y$ positional residuals as a function of $x$ and $y$ raw MIRI positions are shown in the side panels.}
    \label{fig:F560W_GD_01}
\end{figure*}

We start by measuring stellar positions in the MIRI images by fitting our library ePSF models. Only bright, well-measured, unsaturated stars with a \qfit lower than 0.1 were kept in each MIRI catalog. Selections were also applied to the \gaia catalog: we considered in the analysis only stars with $G$ magnitude between 13 (saturation makes \gaia astrometry worse for stars brighter than this threshold) and 19 (to exclude faint stars with poorly-measured proper motions) with a positional error (including the contribution for the proper-motion propagation to epoch mid 2022) lower than 0.01 MIRIM pixel.

We cross-identified the same stars in the MIRI and \gaia catalogs, and transformed the \gaia positions on to the raw reference system of the MIRI imager by means of four-parameter linear transformations (rigid shifts in the two coordinates, one rotation, and one change of scale). Positional residuals were defined as the difference between these transformed \gaia and the raw MIRI positions. We collected all residuals and fit them with two fifth-order polynomial functions. The coefficients of the polynomial functions were obtained via least-square fit of all positional residuals. For the polynomial correction, we chose the center of the imager $(x_{\textrm{ref}},y_{\textrm{ref}}) = (693.5,512.5)$ as the reference pixel with respect to which solve for the GD \citep{2023LibralatoNIRISS}. This pixel is also the reference pixel of the MIRI imager full-frame array used by the Science Instrument Aperture Files (SIAF) of \jwst\footnote{See the corresponding \href{https://jwst-docs.stsci.edu/jwst-observatory-characteristics/jwst-field-of-view}{JDox page} and references therein.}. The GD correction for each star is defined as:
\begin{equation*}
    \left \{
    \begin{split}
        \delta x = \sum_{i = 1, 5}\sum_{j = 1, 5-i} a_{ij} \Tilde{x}^i\Tilde{y}^j \\
        \delta y = \sum_{i = 1, 5}\sum_{j = 1, 5-i} b_{ij} \Tilde{x}^i\Tilde{y}^j
    \end{split}
    \right. \, ,
\end{equation*}
where:
\begin{equation*}
    \left \{
    \begin{array}{c} 
        \Tilde{x} = \frac{x-x_{\textrm{ref}}}{x_{\textrm{ref}}} \\
        \Tilde{y} = \frac{y-y_{\textrm{ref}}}{y_{\textrm{ref}}}
    \end{array}
    \right. \, . 
\end{equation*}
We set to 0 the coefficients $a_0$ and $a_1$ of the polynomial in order to: (i) have the solution at the reference pixel to have its scale equal to that of the chip, and (ii) the corrected $y$ axis aligned to the raw $y$ axis. The terms $b_0$ and $b_1$ were instead left to assume whatever values fit best because the detector axes have different scale and not be perpendicular to each other. For the polynomial part, we performed 500 iterations, each time adding only 75\% of the coefficient values to the previous estimates to ensure a convergence of the solution. Figure~\ref{fig:F560W_GD_01} shows the distortion maps and the positional residuals as a function of $x$/$y$ coordinates before (left) and after (right) applying the polynomial correction in the case of the F560W filter. This first correction improves the positional residuals by a factor $\sim$50, but we can still notice part of the non-linear GD is left uncorrected. The result from the polynomial correction provided in this work is similar to what can be obtained using the GD correction in the corresponding CRDS reference file, which, as well, includes a polynomial correction only.

The residual GD was empirically corrected using a look-up table of residuals. The positional residuals after the polynomial correction for the F560W, F770W and F1000W filters had the same shape and amplitude\footnote{It is worth noticing that the $\Delta y$ as a function of $y$ plot in the rightmost panel of Fig.~\ref{fig:F560W_GD_01} shows a periodic pattern with period of about 374 MIRIM pixels. The pattern is present regardless of the filter, as we would expect for a detector-related feature \citep[e.g.][]{2014LibralatoHAWKI,2023LibralatoNIRISS}.}, and therefore, we chose to work with all of them at once to increase the statistics, at the cost of a less accurate distortion than with a purely filter-based solution. We also relaxed the quality selections on both the MIRI and \gaia catalogs, again to have enough stars to tabulate the correction.

\begin{figure}[t!]
    \centering
    \includegraphics[width=\columnwidth]{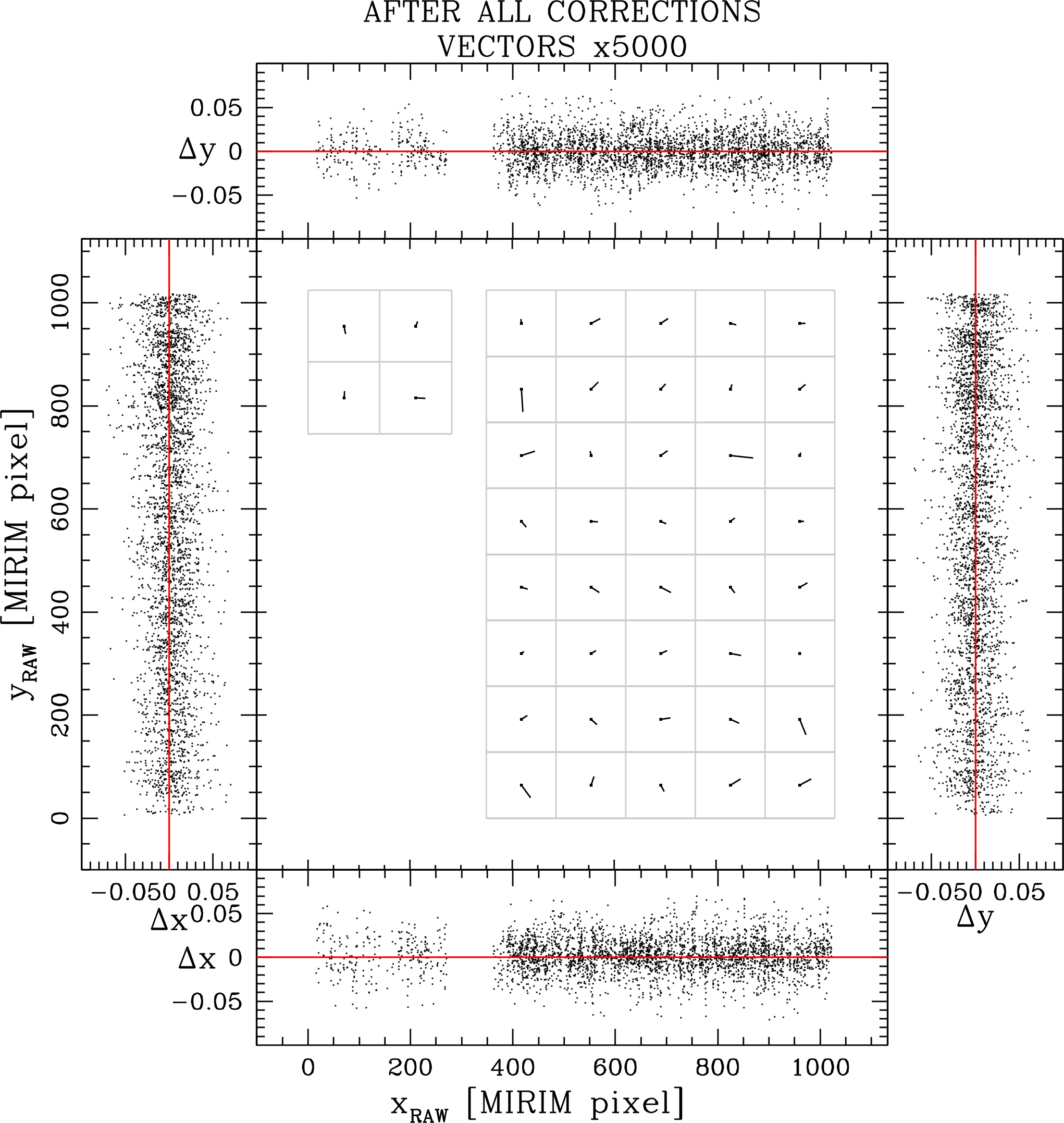}
    \caption{As in Fig.~\ref{fig:F560W_GD_01}, but after the look-up table of residuals correction is also applied. Vectors are now magnified by a factor 5\,000.}
    \label{fig:Filters_GD_2}
\end{figure}

\begin{figure*}[t!]
    \centering
    \includegraphics[width=\textwidth]{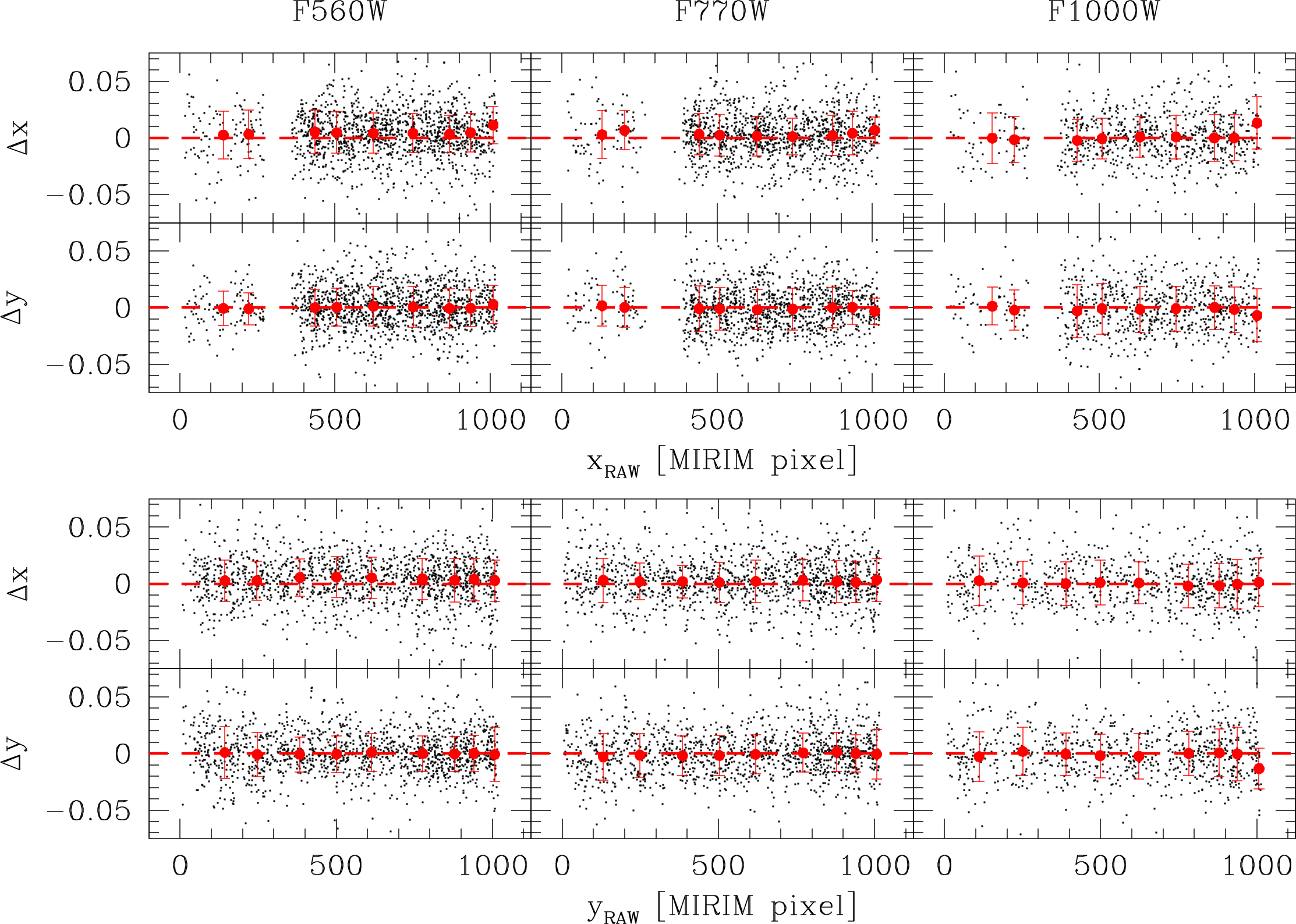}
    \caption{The $x$ and $y$ positional residuals as a function of $x$ and $y$ raw MIRI positions for each of the three filters used to compute the table of residuals correction. The red, dashed horizontal line is set to 0 as a reference. The median and the 68.7-th percentile about the median value of the residuals in 250-pixel-wide bins are shown in red.}
    \label{fig:Filters_GD_3}
\end{figure*}

We collected all positional residuals and averaged them in two look-up tables,  a 3$\times$3 for the Lyot region (141$\times$92.7 pixel$^2$ elements) and a 8$\times$12 (127.9$\times$128.1 pixel$^2$ elements) for the imager. The grid points adjoining to the edges of the Lyot or the imager regions were displaced to edges of the cell, so to allow the correction to be always computed by bi-linearly interpolating between grid elements. Again, the computation of this second correction was iterated 100 times, each time adding 50\% of the correction to ensure convergence. The distortion map and positional residuals after all corrections are applied for all three filters is shown in Fig.~\ref{fig:Filters_GD_2}. Figure~\ref{fig:Filters_GD_3} presents the positional residuals as a function of $x$ and $y$ positions for each filter separately. The residual GD systematics are within 0.01 MIRIM pixel and are larger close to the edges of the detector, where we expect our GD solution to be less accurate because of the lack of sources in the region to model the GD.

\subsection{Astrometric precision and short-term temporal stability}

We evaluated the astrometric precision of our GD correction by combining multiple  dithered MIRI observations from three distinct data sets covering fields in the LMC, specifically PIDs 1024, 1040 and 1521. This allowed us to test our GD using images not related to the GD-computation process and, at the same time, to monitor the short-term stability of the GD since these three sets of observations were taken three months apart.

\begin{figure*}[th!]
    \centering
    \includegraphics[width=\textwidth]{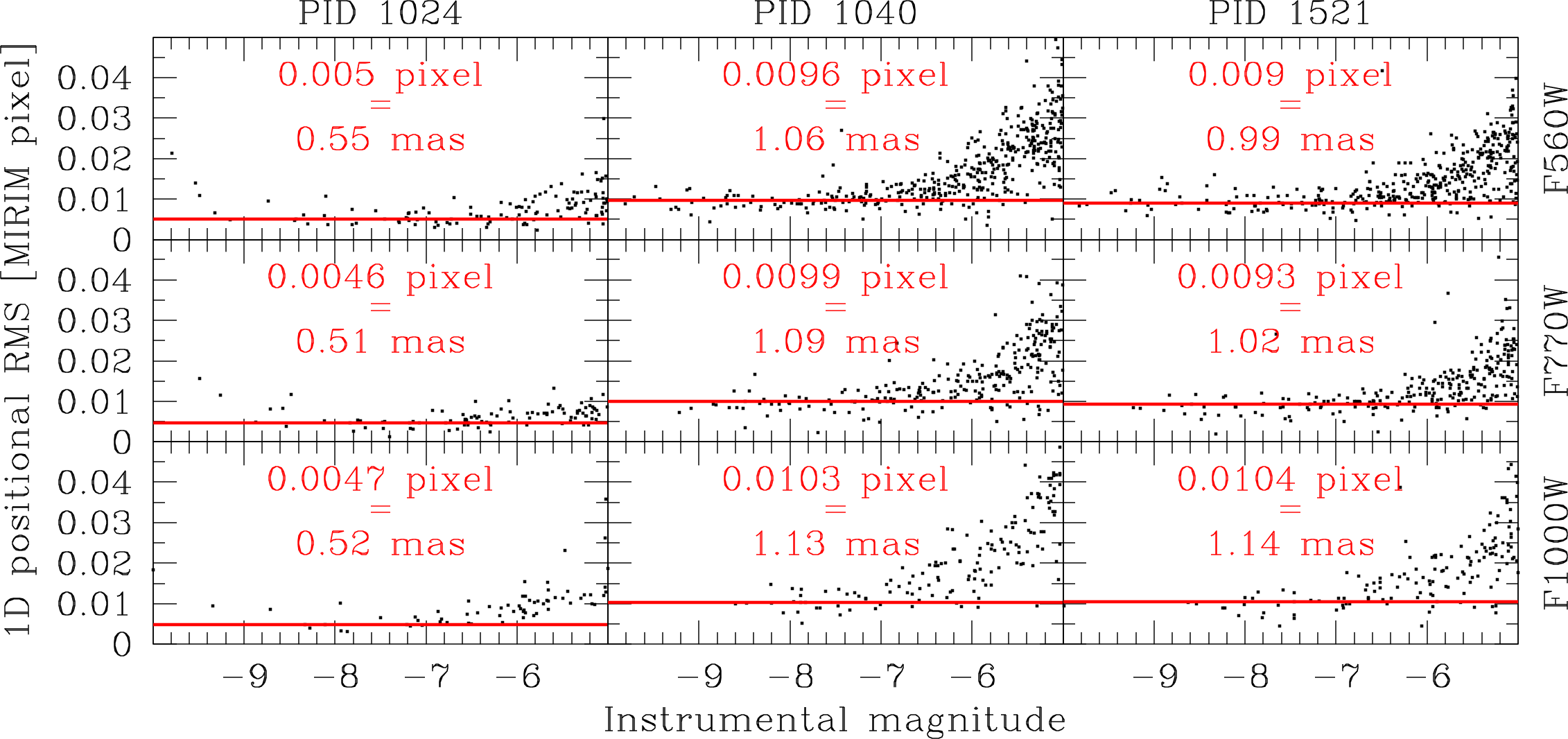}
    \caption{One-dimensional RMS (the sum in quadrature of the positional RMS along the $x$ and $y$ axes divided by $\sqrt{2}$) as a function of instrumental magnitude for the F560W, F770W and F1000W filters from top to bottom, respectively. The first column from the left refers to the results obtained using data from PID 1024, the second from PID 1040 and the third from PID 1521. The red horizontal line is set to the median value of the one-dimensional RMS of bright, unsaturated stars with a \qfit lower than 0.05. The pixel scale used for the pixel-to-mas conversion is the nominal pixel scale of MIRI (110 mas pixel$^{-1}$).}
    \label{fig:1D_RMS}
\end{figure*}

For each program/filter, we made a master frame as described in Sect.~\ref{sec:psf} using six-parameter linear transformations. We only kept stars measured in at least three images. The positional RMS as a function of the instrumental magnitude for all cases is shown in Fig.~\ref{fig:1D_RMS}. The median value of the one-dimensional positional residuals is always of the order of 0.01 pixel or better, in agreement with what found for NIRCam and NIRISS \citep{2023GriggioNIRCam,2023LibralatoNIRISS}. The results for PID 1024 are a factor two better than in the other cases. However, the PID 1024 images were taken with a small dither and the estimate of the astrometric precision with this data set is blind to systematic residuals (a given star falls in the same region the detector in different images, and the GD usually varies smoothly on small scales). Images of PIDs 1040 and 1521 were instead taken in a $3 \times 3$ mosaic with large offsets between tiles and can provide a more reliable measurement of the astrometric precision. Finally, we can also conclude that the non-linear terms of the GD are rather stable over a period of three months.

\subsection{Pixel scale}\label{sec:scale}

We estimated the absolute pixel scale of the MIRI imager by means of the \gaia DR3 catalog. We cross-identified bright and well-measured stars between each MIRI catalog of PID 1521 and the \gaia catalog, and used six-parameter linear transformations to transform positions from one frame to the other. The relative scale between the two frames is one of the linear parameters solved by these transformations. Given that the pixel scale of our master frame was defined to be 110 mas pixel$^{-1}$ (see Sect.~\ref{sec:psf}), we could derive the absolute pixel scale of the MIRI imager from the relative one.

We compute the absolute scale for each filter separately, taking into account for the velocity-aberration correction factor (provided by the header keyword \texttt{VA\_SCALE}). The median values of the pixel scale in each filter analyzed for the GD correction is provided in Table~\ref{tab:scale}. We find that the pixel scale is filter dependent.

\begin{table}[t!]
    \centering
    \caption{Average pixel scale for three MIRI filters (corrected for velocity aberration).}
    \begin{tabular}{c|c|c}
        \hline
        \hline
        Filter & Pixel scale [mas pixel$^{-1}$] & Error [mas pixel$^{-1}$] \\
        \hline
        F560W  & 110.47645 & 0.00029 \\
        F770W  & 110.47894 & 0.00036 \\
        F1000W & 110.40550 & 0.00034 \\
    \hline 
    \hline
    \end{tabular}
    \label{tab:scale}
\end{table}

\begin{figure*}[th!]
    \centering
    \includegraphics[width=\textwidth]{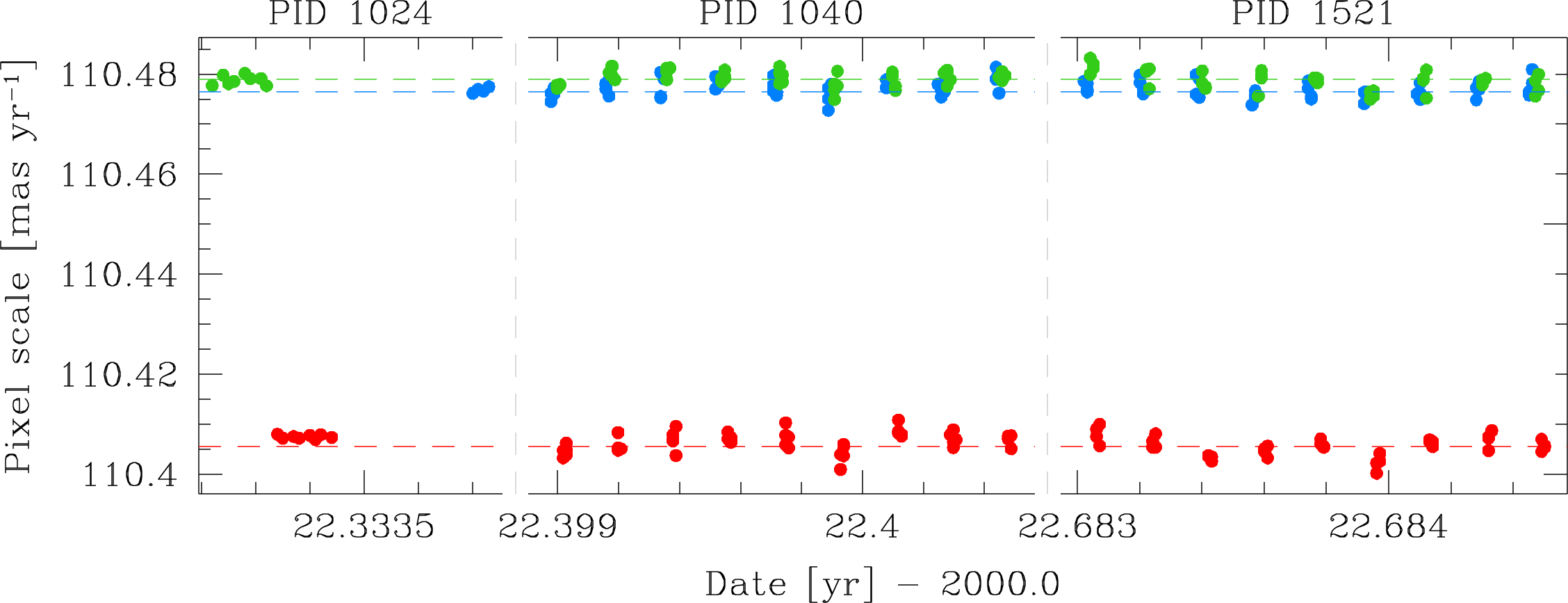}
    \caption{Pixel scale as a function of time. Blue, azure and red points refer to the results from the analysis of F560W, F770W and F1000W filters data, respectively. The dashed, horizontal lines, color-coded as for the points, mark the average pixel scale of PID 1521 data.}
    \label{fig:scale}
\end{figure*}

We analyzed the variation of the pixel scale as a function of time by repeating the same process described before using also data from PID 1024 and PID 1040 (Fig.~\ref{fig:scale}). We find the pixel scale to be rather stable over three months, with variations at the 5$\sigma$ level in the worst case. We also noticed that while the pixel scale is almost constant in all images of PID 1024, the values show an apparent trend as a function of time during the observations of PIDs 1040 and 1521 in all filters. The variations are of the order of $4 \times 10^{-5}$ in $\sim$13 hours of observation. This resembles the pattern of the pixel-scale variations seen in the NIRISS detector by \citet{2023LibralatoNIRISS}. Interestingly, both the NIRISS and MIRI images showing this modulation in the pixel scale have been taken as part of a $3 \times 3$ mosaic with large dithers.

\section{Public release of the MIRI ePSF models and GD corrections}\label{sec:desc}

We release our ePSF models and GD corrections to the community through various channels. We describe here how to use them and provide a few caveats for the users.

We provide two sets of ePSF models\footnote{The MIRI ePSF models are available at \href{https://www.stsci.edu/~jayander/JWST1PASS/LIB/PSFs/STDPSFs/MIRI/}{STDPSFs}.}. In addition to the $301 \times 301$ pixel$^2$ ePSFs described in Sect.~\ref{sec:psf}, we also make available a cutout version with size $101 \times 101$ oversampled pixels$^2$, with the ePSF centered at pixel (51,51). Besides the size and the center, the two sets of ePSFs have the same conventions (the ePSFs are normalized to have unity flux within a radius of 10 MIRI pixels -- 40 oversampled pixels). 

These smaller ePSFs are specifically designed to be used with the FORTRAN routine \texttt{jwst1pass}\footnote{The FORTRAN code is available at \href{https://www.stsci.edu/~jayander/JWST1PASS/CODE/JWST1PASS/}{CODE}.} \citep[Anderson et al., in preparation]{2023LibralatoNIRISS}, which is the \jwst analog of the tool \texttt{hst1pass} \citet{2022acs..rept....2A,2022wfc..rept....5A} designed for ePSF fitting with \hst images. The larger ePSFs are provided to the community for completeness, but can only be used with other tools, at least for now. Various PSF-fitting software packages like \texttt{photutils} \citep{larry_bradley_2022_6825092} can potentially use our ePSFs as far as these tools follow the ePSF conventions and caveats we described in Sect.~\ref{sec:psf}. 

The ePSF file for each filter contains nine ePSF slots, but only eight are filled. Six ($2 \times 3$) ePSFs are reserved for the imager. They can be linearly interpolated to construct the ePSF for a star centered in any pixel of the imager region. The ePSF made for the Lyot region is copied into the center left and upper left slots of the PSF array, which allows the PSF to be constant within the Lyot region. This formulation allows \texttt{jwst1pass} to use the generic STDPSF-based software developed for all \hst and \jwst images with MIRI.  Specific header keywords are provided in each file with the fiducial detector location for each of the nine PSFs \citep[see][]{2022acs..rept....2A,2022wfc..rept....5A}. The ePSFs for the 4QPMs are not provided since this region is not calibrated for  direct (as opposed to coronagraphic) imaging. 

The tool \texttt{jwst1pass} uses the centermost $5 \times 5$ pixel$^2$ of a star to fit the ePSF model. We have noticed that at long wavelengths (F2100W and F2550W filters) faint stars are not always detected. These sources do not have a well-defined local maximum because of the high pixel-to-pixel noise. Finally, we stress again that the ePSFs at these wavelengths, especially for the F2550W filter, were obtained with few stars, and they are less accurate. For all these reasons, we advise users to carefully interpret the result of the ePSF fit and evaluate it on a case-by-case basis using the various diagnostic parameters and images output by \texttt{jwst1pass}.

The GD corrections are released as data-cube FITS file\footnote{The MIRI GD corrections are available at \href{https://www.stsci.edu/~jayander/JWST1PASS/LIB/GDCs/STDGDCs/MIRI/}{STDGDCs}.} designed to work with the \texttt{jwst1pass} software. We refer to \citet{2023LibralatoNIRISS} and Appendix~G of \citet{2022acs..rept....2A,2022wfc..rept....5A} for the description of the GD format. We also release\footnote{\href{https://github.com/mlibralato/MIRIGDC}{https://github.com/mlibralato/MIRIGDC}.} a \texttt{python} tool that applies our GD corrections to a list of coordinates. Note that positions must be defined in a 1-index reference frame. Positions in a 0-index \texttt{python}-like frame would need to offset by one pixel in each coordinate before applying the GD correction.

\section{Example of scientific application}\label{sec:science}

We tested our tools by studying a field in the LMC observed as part of the Commissioning PID 1171 in F560W and F770W filters. We chose this data set because it was employed neither for the ePSF modeling nor for the GD correction, and it thus offers us an independent benchmark.

\begin{figure}[t!]
    \centering
    \includegraphics[width=\columnwidth]{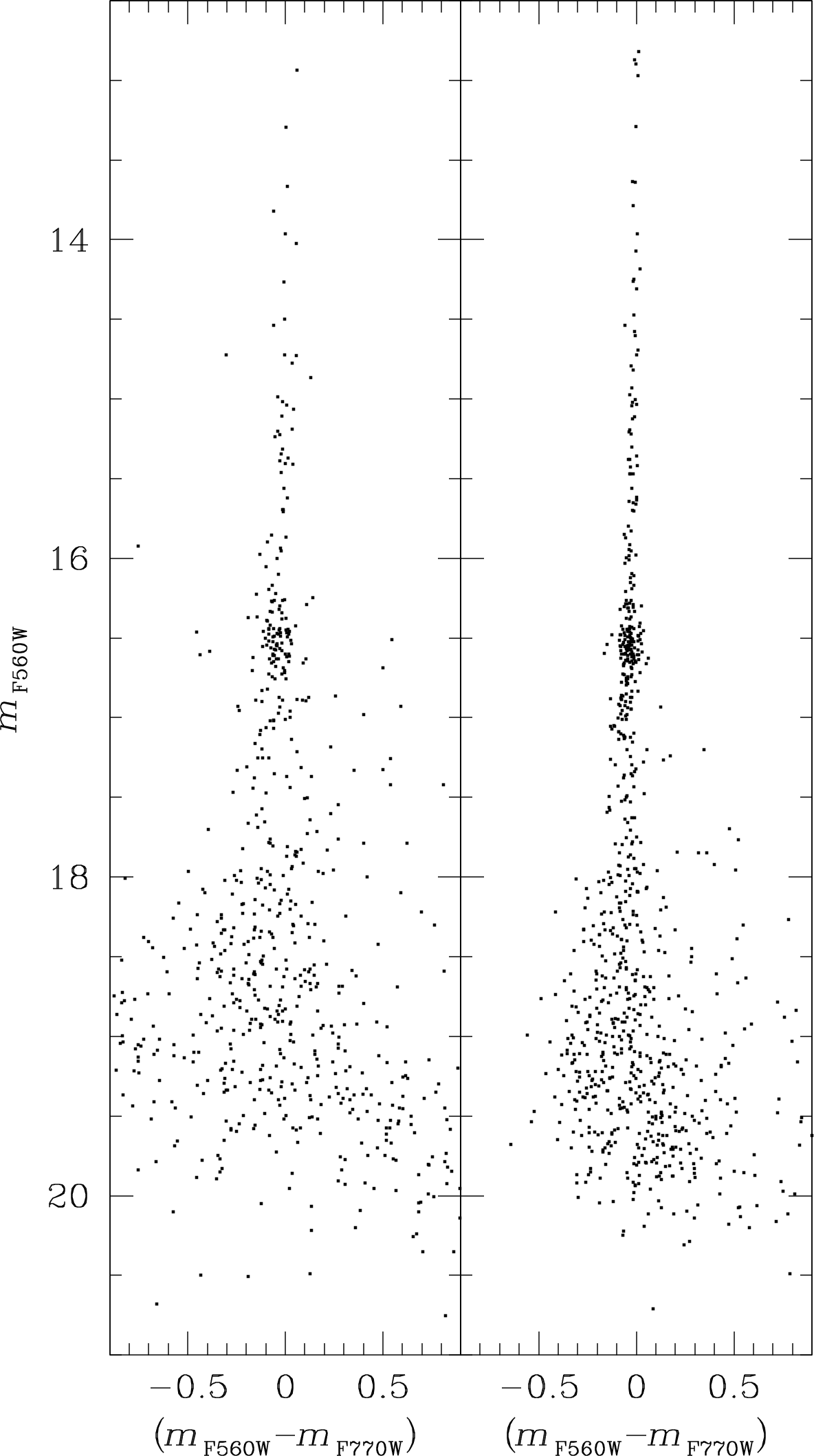}
    \caption{Calibrated $m_{\rm F560W}$ versus $(m_{\rm F560W}-m_{\rm F770W})$ CMD for the stars in the LMC field imaged by PID 1171. The left panel shows the result obtained with aperture photometry on the LVL-3 mosaic image with the current \jwst pipeline, while the right panel refers to the ePSF-based photometry on the LVL-2 images.}
    \label{fig:cmd}
\end{figure}

\begin{figure*}[th!]
    \centering
    \includegraphics[width=\textwidth]{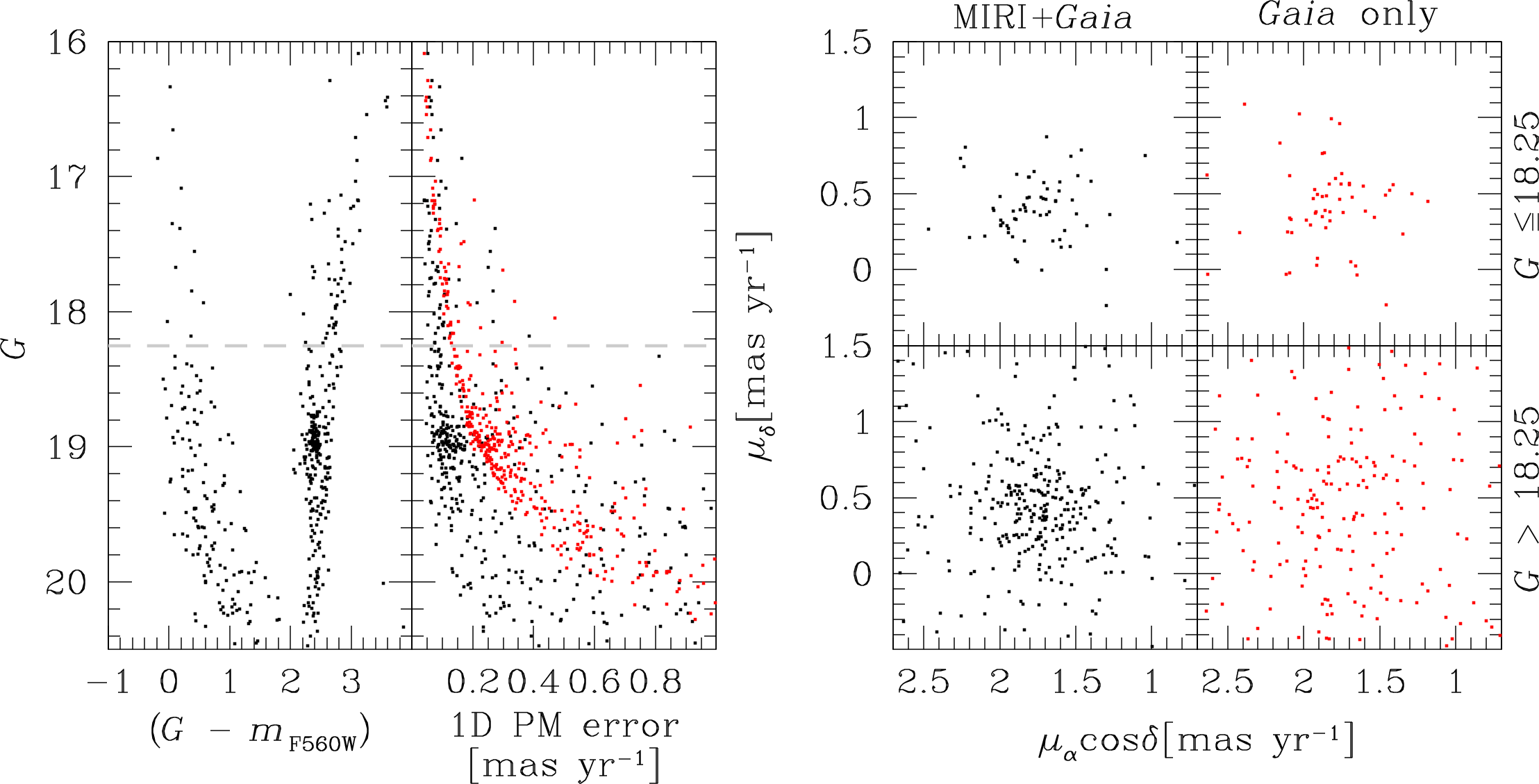}
    \caption{Overview of our PM test using the MIRI data of PID 1171 and the \gaia catalog. From left to right, we show $G$ versus $(G-m_{\rm F560W})$ color-magnitude diagram (CMD); the PM error as a function of $G$; the vector-point diagram (VPD) of the stars with our PMs; the VPD of the same stars with \gaia's PMs. In all but the leftmost panel, black points refer to our PM catalog, whereas the red points marks the analog quantities in the \gaia catalog.}
    \label{fig:pm}
\end{figure*}

We fitted all sources in these images using our ePSF models. Positions were then corrected for GD using our corrections. For each filter, we made a master frame as described in Sect.~\ref{sec:psf}, using the \gaia catalog to setup orientation and scale. Finally, we identified the stars in common between the F560W and F770W star lists. The color-magnitude diagram (CMD) of the stars in this LMC field is shown in the right panel of Fig.~\ref{fig:cmd}. The left panel presents the CMD obtained using aperture photometry on the LVL-3 mosaic image with the current \jwst pipeline. Our photometry was registered on to the VEGA-mag system by computing the zero-point between our photometry and that of the pipeline using bright, well-measured stars in common. PSF and aperture photometry should be comparable in this field because it is not very crowded. However, the MIRI ePSF photometry provides a narrower sequence in the CMD at all magnitudes, not only at the faint end as expected \citep{2016LibralatoK2i,2016LibralatoK2ii}.

The accuracy of our GD correction was tested by computing proper motions (PMs) for the stars in this field. We cross-identified the stars in our MIRI F560W catalog with those present in the \gaia DR3 catalog and transformed (with global, six-parameter linear transformations) the positions of the stars in the MIRI master frame on to the reference frame of the \gaia catalog. We used only stars belonging to the old LMC population along the red-giant branch in the CMD (left panel of Fig.~\ref{fig:pm}) to compute the coefficients of the transformations between frames. Thus, our PMs are initially relative to the bulk motion of these old stars. PMs are defined as the positional displacements divided by the temporal baseline ($\sim$8 years) and multiplied by the pixel scale of the \gaia catalog (110 mas pixel$^{-1}$). PM errors are computed as the sum in quadrature of the positional errors, again divided by the temporal baseline and multiplied by the pixel scale. Finally, we linked our relative PMs on to the absolute frame of \gaia by computing the PM zero-point between our and \gaia's PMs. We provide the $G$ versus $(G-m_{\rm F560W})$ CMD and the PM error as a function of $G$ in Fig.~\ref{fig:pm}. The comparisons between the vector-point diagram (VPD) of the stars in this LMC field using our (black points) and \gaia's (red points) PMs in two magnitude regimes are also shown. For the brightest stars in common between the two catalogs, the MIRI$+$\gaia PMs are comparable, with a median PM error of $\sim$80 $\mu$as yr$^{-1}$. At the faint end, our PMs are more precise because of the long temporal baseline and worse quality of the \gaia PMs. It is worth noticing that these results were obtained with simple techniques, and we expect a more careful analysis \citep[e.g.,][]{2022LibralatoPMcat} to provide more precise PMs.

\section{Conclusions}\label{sec:conc}

We presented a careful analysis of the astro-photometric capabilities of the MIRI imager. We described in detail how accurate ePSF models and GD corrections were made, and explain how to use them to obtain PSF-based photometry with \jwst data, which is currently not performed by any step of the \jwst imager pipeline.

We made ePSF models for all MIRI filters using public Commissioning, Cycle-1 Calibration and GO data. We refer the reader to the discussions throughout Sects.~\ref{sec:psf} and \ref{sec:desc} for how to use these models and for the associated caveats. We analyzed the impact of the brighter-fatter effect in MIRI PSF photometry. This, combined with other effects like non-linearity, can impact the photometry at a few-percent level and result in imperfect PSF subtraction. For the GD, we instead release only the corrections of the three shortest-wavelength filters among those available for MIRI. We show that GD is stable, at least in the short timescale analyzed in our work.

We tested our ePSFs and GD corrections by analyzing a stellar field in the LMC. We showed that PSF photometry outperforms what can be obtained using aperture photometry. We also computed PMs by combining \jwst MIRI and \gaia catalogs, which can be a powerful combination to analyze the kinematics of stars in the faint regime of \gaia \citep[e.g.,][]{2022delPinoGaiaHub}.

These simple applications highlight the astrometric and photometric potential of the MIRI imager. The complete assessment of the astro-photometric capabilities of this imager is still to be completely understood; only more data spanning a large variety of exposure parameters and the impact of the pipeline and calibration will enable us to learn more about this instrument. The MIRI resolution is worse than that of the NIR \jwst detectors but the wavelength coverage of MIRI is unique among \jwst's detectors, making the MIRI imager an invaluable tool at disposal of the astronomical community.

\section*{Acknowledgments}

ML thanks Marshall Perrin, Marcio Melendez Hernandez and Andrea Bellini for the insightful discussions about WebbPSF, ePSFs and GD. ML also thanks Jay Anderson for implementing the MIRI module in \texttt{jwst1pass} and for hosting the ePSF and GD-correction files on his web page. IA would like to thank the European Space Agency (ESA) and the Belgian Federal Science Policy Office (BELSPO) for their support in the framework of the PRODEX Programme. PJK acknowledge support from Science Foundation Ireland/Irish Research
Council Pathway programme under Grant Number 21/PATH-S/9360. Based on observations with the NASA/ESA/CSA \textit{JWST}, obtained at the Space Telescope Science Institute, which is operated by AURA, Inc., under NASA contract NAS 5-03127. This work has made use of data from the European Space Agency (ESA) mission {\it Gaia} (\url{https://www.cosmos.esa.int/gaia}), processed by the {\it Gaia} Data Processing and Analysis Consortium (DPAC, \url{https://www.cosmos.esa.int/web/gaia/dpac/consortium}). Funding for the DPAC has been provided by national institutions, in particular the institutions participating in the {\it Gaia} Multilateral Agreement. This research made use of \texttt{astropy}, a community-developed core \texttt{python} package for Astronomy \citep{astropy:2013, astropy:2018}.

\facilities{\jwst (MIRI), \gaia}
\software{FORTRAN, \texttt{python}, \texttt{astropy}, \texttt{photutils}}

\clearpage
\appendix

\setcounter{table}{0}
\renewcommand{\thetable}{A\arabic{table}}

\setcounter{figure}{0}
\renewcommand{\thefigure}{A\arabic{figure}}

\begin{figure*}[th!]
    \centering
    \includegraphics[width=\textwidth]{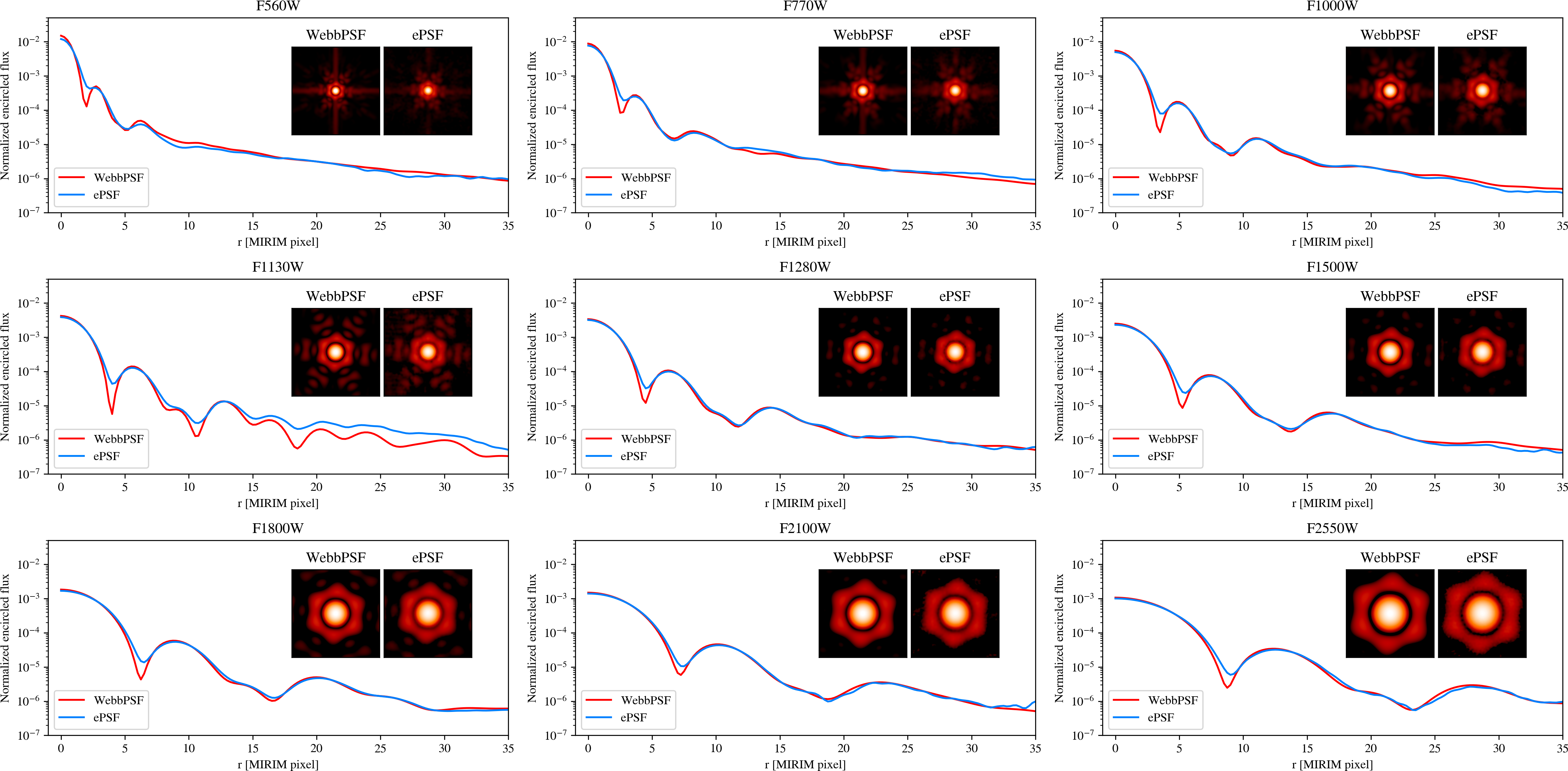}
    \caption{Comparison between the normalized encircled-flux radial profiles of WebbPSF (red) and our ePSF (blue) models. The insets in each panel show a $75 \times 75$ oversampled pixel$^2$ zoom-in around the center of each oversampled model.}
    \label{fig:all_comp}
\end{figure*}

\section{Comparison with WebbPSF models}\label{sec:webbpsf}

\begin{figure*}[th!]
    \centering
    \includegraphics[width=\textwidth]{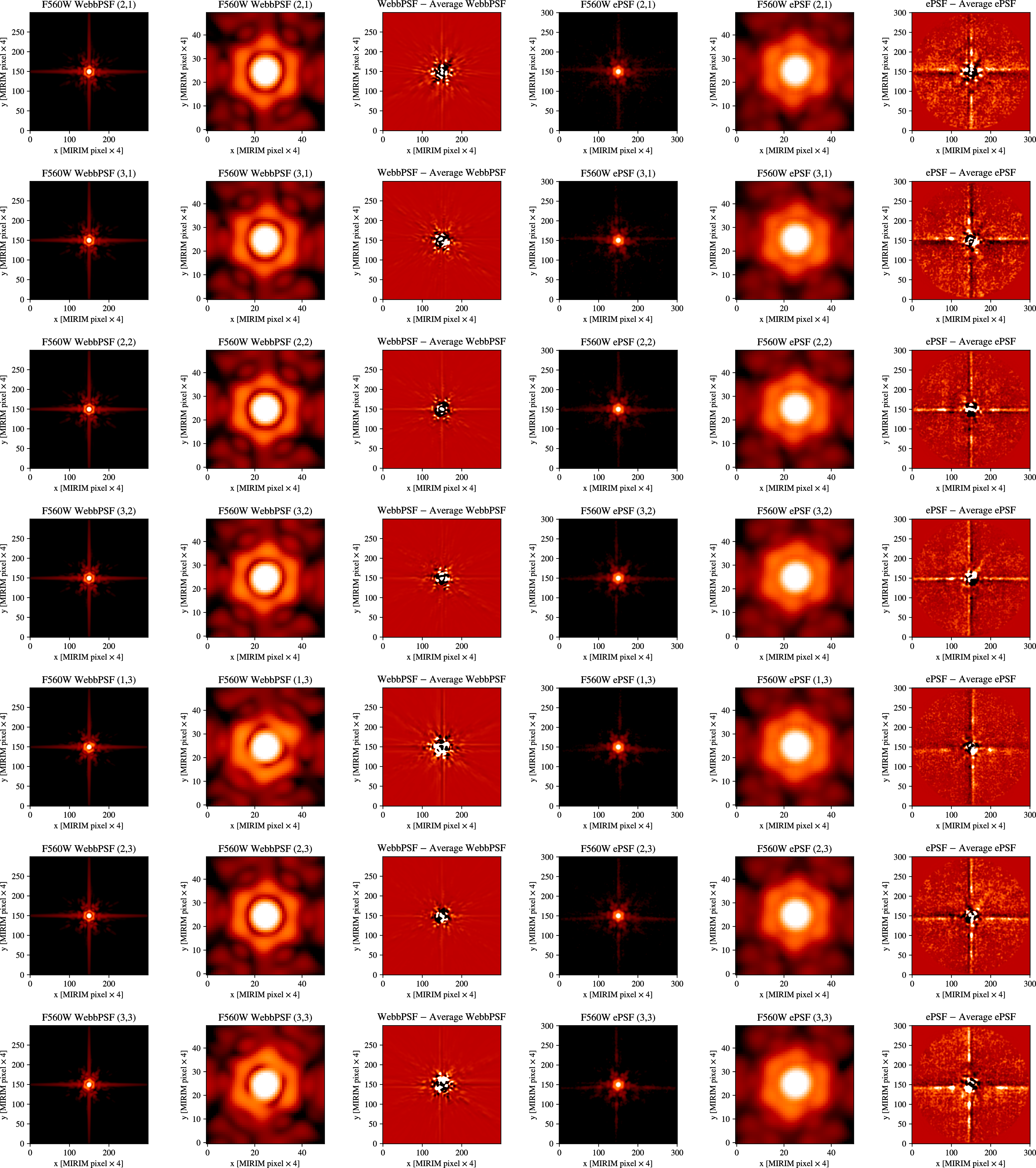}
    \caption{Comparison between the WebbPSF and ePSF models for the F560W filters. The three leftmost panels show, respectively, the WebbPSF PSF, a zoom-in in its core (both in log scale) and the difference with respect to the average PSF in the field of view (in linear scale). The rightmost three panels present analog plots using our ePSFs (with the same scales as the corresponding WebbPSF panel). The naming convention follow the scheme shown in Fig.~\ref{fig:ePSF_scheme}.}
    \label{fig:f560w_comp}
\end{figure*}

Simulated PSFs for the \jwst instruments are available thanks to the \texttt{python} package WebbPSF. In the version used for our test (1.1.2, developer version), WebbPSF PSFs takes into account models of the telescope and optical state of an instrument but does not include detector effects like interpixel capacitance and intrapixel sensitivity variations. The only detector effect included for MIRI is the cruciform artifact, although the simulated cruciform properties are significantly different from what is observed in flight data (see below). This qualitative comparison is meant to highlight the importance of the detector effects.

We simulated one PSF with WebbPSF for each filter and compared it with one of our ePSFs. For filters where the spatial variability was included in the modeling, we considered the centemost ePSF in our array. Figure~\ref{fig:all_comp} presents the encircled-flux profiles obtained from the WebbPSF (red lines) and our ePSF (blue lines) models. These profiles are normalized to have unity flux within a radius of 10 MIRIM pixels. Insets in each panel show the centermost part of both models. These profiles and insets suggest that both models are rather similar, although the first minimum of our ePSFs predict more flux than WebbPSF. This is likely due to the smoothing functions used in the ePSF modeling (see Sect.~\ref{sec:psf}), and to the lack of detector effects in WebbPSF PSFs. The F1130W ePSFs predict more flux than those of WebbPSF at all radii. Our F1130W ePSFs were obtained from data where stars are embedded in dust, which caused a gradient in the wings of the ePSFs in this filter. We choose to include the wings anyway, but we advise caution when using them.

The F560W and F770W PSFs include the cruciform artifact. As a benchmark, we focused on the F560W filter and simulated seven PSFs located at the same reference points of our ePSF models. We provide a detailed comparison for the F560W case in Fig.~\ref{fig:f560w_comp}. The three leftmost panels show, respectively, the WebbPSF PSF, a zoom-in in its core and the difference with respect to the average PSF in the field of view. The rightmost three panels present analogous plots using our ePSFs.

The core of both sets of PSFs is rather similar\footnote{The internal measurements of the wave-front errors from the Integrated Science Instrument Module (ISIM) testing campaign CV3 did not cover the top-left (1,3) and top-right (3,3) corners of the MIRI imager. For this reason, WebbPSF models had to be extrapolated, and this creates the asymmetric feature clearly visible in the corresponding WebbPSF models (see the \href{https://webbpsf.readthedocs.io/en/latest/jwst.html\#id12}{WebbPSF documentation}).}. The main differences are present in the wings of the PSF, where the cruciform artifact is prominent. WebbPSF PSFs seem to predict more flux in the cruciform artifact than our ePSFs that does not vary significantly across the field of view. Instead, our ePSFs show that the cruciform artifact is weaker than in the WebbPSF models and a more evident spatial dependence, especially for the horizontal component. Also, the cruciform seems to be bent \citep{2021GasparCruciform}.

In light of these pieces of evidence, we advise caution when using WebbPSF models, at least for now. Despite providing an accurate description of \jwst and its instruments' optical model, systematic effects might prevent users from achieving high-precision astrometry and photometry with WebbPSF PSFs, because detector effects like interpixel capacitance, intrapixel sensitivity and, for the F560W and F770W filters, the cruciform artifacts is not accurately modelled. Preliminary tests made with the latest WebbPSF version (1.2.1), which includes the effect of the interpixel capacitance of MIRI (Engesser et al., in preparation) and a cruciform model closer to that of flight data, show a better agreement with our ePSF models.

\bibliography{MIRI_astro.v2}{}
\bibliographystyle{aasjournal}

\end{document}